\documentclass[11pt]{article}

\usepackage{latexsym, amssymb, enumerate, amsmath}
\usepackage{lscape}
\usepackage{amsfonts}
\usepackage{fancyhdr}
\usepackage{setspace}
\usepackage{color}
\usepackage{enumitem}
\usepackage{longtable}
\usepackage{float}
\usepackage{graphicx}
\usepackage{caption}
\usepackage{subfigure}
\usepackage{hyperref}
\usepackage{natbib}
\usepackage{booktabs}
\renewcommand\footnotemark{}

\newcommand{\bfm}[1]{\mathbf{#1}}

\newcommand{\ds}{\displaystyle}

\newcommand{\be}{\begin{equation}}
\newcommand{\ee}{\end{equation}}
\newcommand{\bdm}{\begin{displaymath}}
\newcommand{\edm}{\end{displaymath}}
\newcommand{\bc}{\begin{center}}
\newcommand{\ec}{\end{center}}

\date{}
\title{
\textbf{A discrete in continuous mathematical model of cardiac progenitor
cells formation and growth as spheroid clusters (Cardiospheres)}
}
\author{{\em Ezio Di Costanzo$^{1*}$,  Alessandro Giacomello$^{2**}$, Elisa Messina$^{3**}$, Roberto Natalini$^{4*}$},\vspace{1mm}\\ {\em Giuseppe Pontrelli$^{5*}$, Fabrizio Rossi$^{6**}$,  Robert Smits$^{7***}$, Monika Twarogowska$^{8*}$}
\footnote{E-mail: $^1${\tt ezio.dicostanzo@gmail.com}, $^2${\tt alessandro.giacomello@uniroma1.it},}
\footnote{$^3${\tt elisa.messina@uniroma1.it}, $^4${\tt roberto.natalini@cnr.it}, $^5${\tt giuseppe.pontrelli@gmail.com},}
\footnote{$^6${\tt rossifabrizio87@gmail.com}, $^7${\tt rsmits@nmsu.edu}, $^8${\tt mtwarogowska@gmail.com}}
\vspace{3mm}\\
$^*$Istituto per le Applicazioni del Calcolo ``M. Picone''\\ Consiglio Nazionale delle Ricerche\\
Via dei Taurini 19 -- 00185 Rome, Italy.
\vspace{3mm}\\
$^{**}$Department of Molecular Medicine \\
Pasteur Institute Cenci--Bolognetti 
Foundation \\  Sapienza University of Rome, Viale del Policlinico -- 00161 Rome, Italy.
\vspace{3mm}\\
$^{***}$Department of Mathematical Sciences \\
New Mexico State University \\
Las Cruces, New Mexico, USA 88003.
}
\begin{document}
\maketitle

\begin{abstract}
We propose a \emph{discrete in continuous} mathematical model describing the in vitro growth process of biophsy-derived mammalian cardiac progenitor cells growing as clusters in the form of spheres (\emph{Cardiospheres}). The approach is hybrid: discrete at cellular scale and continuous at molecular level. In the present model cells are subject to the self-organizing collective dynamics mechanism and, additionally, they can proliferate and differentiate, also depending on \emph{stochastic processes}. The two latter processes are triggered and regulated by chemical signals present in the environment. Numerical simulations show the structure and the development of the clustered progenitors and are in a good agreement with the results obtained from in vitro experiments.
\end{abstract}
\bigskip
\\\textbf{Mathematics Subject Classiﬁcation.} 92B05, 92C17, 92C15.

\section{Introduction}

First described in 2004, the cardiac biopsy-derived progenitor cells, growing in vitro as ``niche''-like microtissue, known as {\em Cardiospheres} (CSps), represent a widely adopted platform technology holding great promise to realize a powerful cell therapy system, in vitro drug screening and disease modeling opening new opportunities for myocardial repair \citep{mess,circul,cheng}. To achieve this goal, an optimal management of qualitative and quantitative CSps growth-modifying factors in terms of cell proliferation, differentiation and interaction with the external environment has to be known. It requires a simulator of the features and the fate of the Cardiospheres-niche-system which is based on new as well as already existing experimental biological data and on reliable mathematical models. 

\begin{figure}
\centering
\subfigure[]{\includegraphics[width=0.5\textwidth]{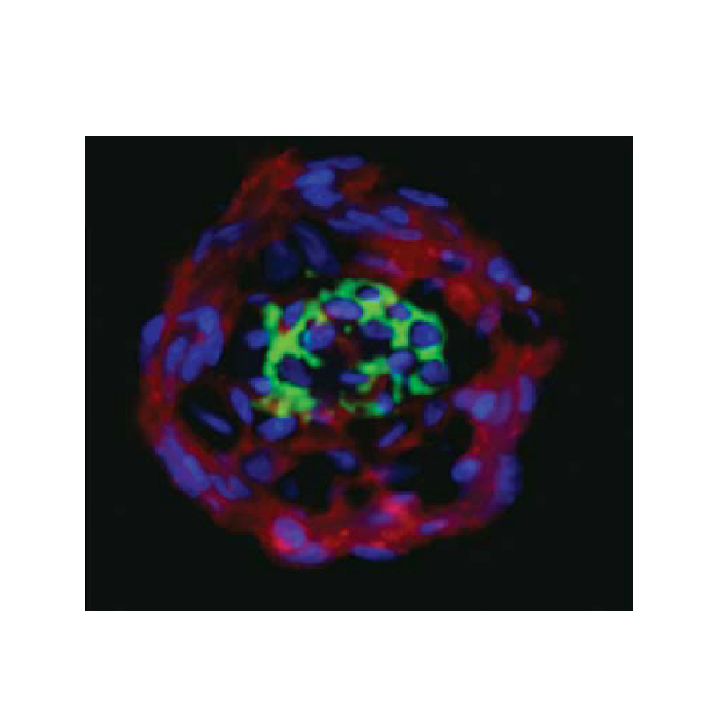}\label{fig1_a}}\hspace{-1.5 cm}
\subfigure[]{\includegraphics[width=0.5\textwidth]{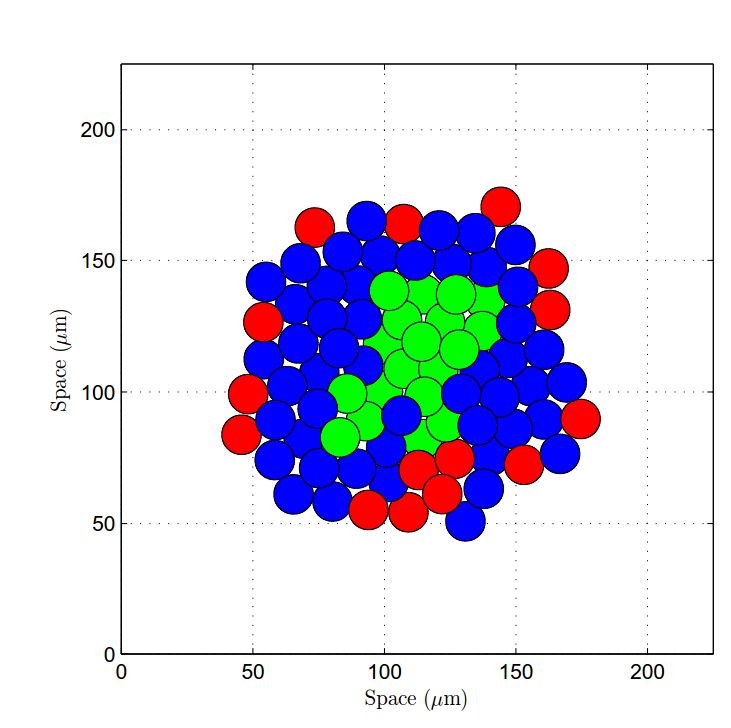}\label{fig1_b}}
\caption{(a) A CSp by immunofluorescence after 3 days: a cluster of immature stem cells (green) is localized in the central core of Cardiospheres, surrounded by differentiated supporting cells (red). Blue spots indicate cell nuclei, and cells inside the black area (non labeled) are to be intended in an intermediate level of differentiation, see also Section \ref{sec:problem} (source from \citealp{cheng}, by courtesy of John Wiley \& Sons). (b) An example of a numerical simulation of our mathematical model showing a CSp at 3 days. Green, blue and red colors mark respectively three increasing degrees of maturation of a single cell (see Section \ref{sec:numeric} for further details).}   
\label{fig1} 
\end{figure}

\begin{figure}
\centering\includegraphics[width=0.5\textwidth, angle=0]{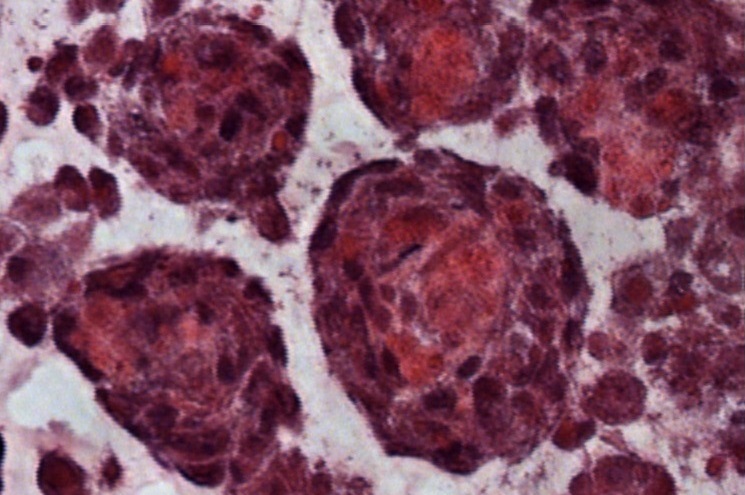}  
\caption{Histological preparation of CSps ultra-thin slices. Hemathoxilin-eosin staining shows a nuclei-free central necrotic area more evident in larger CSps. 
}
\label{fig2}
\end{figure}

Our aim is to model the formation and differentiation of the original 3D cardiac biopsy-derived heart progenitors in vitro system, that is the CSp. These structures are cellular spheroids with a central nucleus of less differentiated elements surrounded by outer layers of cells more committed toward different levels of cardio-vascular differentiation with other proliferation features (Figure \ref{fig1_a}). Their individual formation in vitro starts from a few (low differentiated and less specialized) progenitor cells which are cultured in a carefully chosen environment and under specific conditions \citep{mess,chim1}, and have the potential to develop into a more cardiac-specific differentiation which, unless  derived from pre-natal hearts, are no longer able to achieve a terminal level of contractile cardiomyocytes.
 The major types of molecular processes that control cellular differentiation and proliferation involve various physical factors, nutrients and cell signaling.  During the multiple stages of development, the cell size and shape changes dramatically, as a response to signaling molecules \citep{brown,agath}. 
This biological environment is similar to that of  tumor spheroids as described in \citet{wall}. Neoplastic cells hold several growth strategies, such as genetic mutation of cell cycle inhibitors, neoangiogenesis, etc., and cannot be directly compared with those more limited of the adult cardiac progenitors. Nevertheless both of the 3D cellular growth mechanisms share some common features. CSps and neoplastic spheroids hold more than two functional phenotypes (proliferating, quiescent, differentiating, dying). The growth in a form of a 3D sphere limits the diffusion of the nutrients and oxygen and leads to the formation of a central core of death cells as a consequence of necrosis, apoptosis and anoichisis (Figure \ref{fig2}). It is well known that in tumors the fastest proliferating cells are present in the intermediate and external layers forming the so-called growing front. On the contrary, in small and medium CSps the highest rate of the proliferation is observed in the central area due to the hypoxic conditions, which are closer to those corresponding to the pre-natal developing heart. 
 However, while the spheres become larger the necrotic core appears as in tumors.
%
%
%
In this paper we introduce a discrete in continuous non-deterministic model to simulate the growth of a CSp aggregate. We treat the various biological constituents using different levels of description: on the cellular scale we adopt a discrete description, while on the molecular scale chemical signals are considered as continuous variables \citep{anderson,preziosi}. This choice is justified by the relatively small number of cells involved in the formation of CSps, for which the hypothesis of the continuum does not hold, and by the much smaller size of the molecules with respect to them. Similar approaches have been adopted in analogous cell systems, for instance in relation to the modeling of tumor growth (\citealp{enderling2012,kim,osborne2010,Chaplain2011,fletcher2015}, and references therein) or, recently, to describe the morphogenesis of the lateral line in the zebrafish embryo \citep{dicostanzo1}, and to study the behavior of the collective motion of cells under chemotaxis and alignment effect \citep{dicostanzo2}. With particular regard to the stem cell cultures, in \citet{wu} a continuous diffusion--reaction mathematical model is proposed to investigate oxygen transport and distribution in embryonic stem cell aggregates, also in comparison with experimental data. 
  
Our proposed mathematical model relies on a hybrid description: each cell is described separately by a system of ordinary differential equations, which take into account the self--organizing interactions between cells, through typical terms of collective dynamics and chemotaxis effects (see for instance the reviews by \citealp{vicsek-collective,vicsek-collective-cell}); at a macroscopic level, by reaction-diffusion equations, we describe the concentrations of oxygen, as a representative of nutrients essential for cells metabolism, and of a growth factor TGF-$\beta$ (Transforming-Growth-factor-$\beta$), as a chemoattractant and main regulator of the cell differentiation, secreted by the cells and also present in the environment. Additionally, to model the formation of the Cardiosphere, we introduce two processes, the cell proliferation and differentiation, which lead to the growth and maturation of an initial cell cluster. Moreover, we distinguish three differentiation levels of cell maturation, which also influence the two aforesaid processes.
Moreover they are regulated by the local concentrations of oxygen and TGF-$\beta$. Another crucial feature of our model is the stochasticity. It is motivated by the so called noise in biological systems and the averaging of cells characteristics. The former is mainly due to the genetic diversity among individual cells, random collisions and thermal fluctuations in chemical reactions and randomness of various external factors, see \citet{Tsimring}. The latter concerns the simplifications introduced in our model, such as the uniform size of cells, equal maturation time or signal sensing. The randomness of the biological phenomena in our model is present in the proliferation and differentiation processes. When the necessary conditions are satisfied cells divide or pass to a higher differentiated level with some probability, which increases if the conditions are favorable.            

Using the proposed mathematical model we have simulated the formation of a Cardiosphere in a two--dimensional setting. After a careful sensitivity analysis of the model parameters we were able to obtain numerical results comparable with the existing biological experiment: see again Figure \ref{fig1}, in which the structure and composition of the two images, real experiment (a) and numerical simulation (b), are in good agreement. In order to meet the need of biological research and possible therapeutic applications (including the achievement of a more advanced differentiation level) we have analyzed the development of the Cardiosphere at two typical experimental concentration of oxygen.

The paper is organized as follows: in Section~\ref{sec:problem} we present the general setting of the problem and the mathematical framework used throughout the paper. In Section~\ref{sec:model} we describe in detail the mathematical model for the cells and nutrients dynamics, together with the stochastic mechanisms of cell proliferation and differentiation. Section~\ref{sec:numeric} contains the numerical results and relative biological interpretations. Finally, Section \ref{conclusion} is devoted to the conclusions and possible future perspectives.

\section{Formulation of the problem}\label{sec:problem}

The in vitro 3D CSp formation starts from a low-density seeding of the explants-biopsy-derived cells in a defined culture conditions. First, 
adhered to the culture dish bottom the individual  stem cells start to 
aggregate in small clusters of very few elements. Reaching certain dimensions they detach and grow immersed in a liquid environment and in a controlled gas atmosphere, which contain substances essential for their metabolism. 
In our model we have reduced to two the biological and physical variables affecting the CSps growth and differentiation: the oxygen, representing nutrients and gases, and the TGF-$\beta$ which is present in the culture medium and is produced by the cells themselves. The latter represents a key modulator of the main pathway involved in the induction of the non-terminal differentiation level achieved in our basic experimental conditions \citep{forte}. 
We assume that during the formation of a CSp, the cells are subject to two basic cellular processes, the proliferation and the differentiation, 
which depend, with some probability, on the type of a cell and on the local availability of oxygen. Moreover, we consider three different increasing differentiation levels for living cells and, separately, dead cells. The complex differentiation mechanism is triggered by a large enough 
concentration of TGF-$\beta$,  is hindered by the presence in the closest proximity of a large number of 
cells of the same type 
and is inhibited when the concentration of oxygen 
drops below a threshold value (see Section \ref{sec:prol_diff}). 

To describe the formation of CSps in the next section we propose a hybrid model, in which cells are 
considered as discrete entities while oxygen and TGF-$\beta$ concentrations are continuum variables. We assume that in the absence of specific signals the cell dynamics is due to attractive-repulsive forces and friction effect, taking place isotropically. The TGF-$\beta$, produced by the cells and also present in the external medium, is a chemoattractant for cell which moves towards its higher concentrations. For computational simplicity a two-dimensional description is adopted. 
 

\subsection{Mathematical notations and operators}\label{sec:mathematical_model}

\begin{figure}
\centering\includegraphics[width=0.45\textwidth]{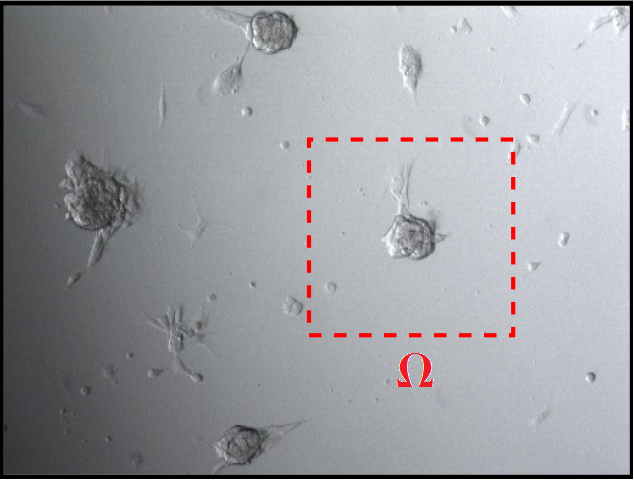}  
\caption{Experimental picture of a culture of CSps. The dotted red line indicates the portion chosen as possible domain $\Omega$ for a single CSp in our mathematical model.}
\label{fig3}
\end{figure}

In a Cartesian coordinate system, let us consider a rectangular domain $\Omega$, which in our model represents a limited portion of the culture (containing many CSps) with a single cluster of stem cells in its center (Figure \ref{fig3}). For this reason in the following we can assume periodic boundary condition in relation to the equations of the model.
At the initial time, $N_0$  cells at the same maturation level (Section \ref{sec:prol_diff}) and radius $R$ (for simplicity all cells are assumed to have the same constant radius) are placed randomly at positions $\bfm X_i, \, i=1,2,...,N_0$ at the center of the domain. In the following, due to proliferation, the number of cells at the time $t$ will be denoted by $N(t)$. In the model we denote by $c(\mathbf{x},t)$ and $S(\mathbf{x},t)$ the concentrations of oxygen and TGF-$\beta$ respectively, and in the numerical simulations these variables are discretized on a uniform spatial mesh. In general,  the cell position $\bfm X_i$ does not coincide with a mesh grid point,  and we have to ``approximate'' the value of $c(\bfm X_{i}, t)$. 
In general, a cell ''feels'' the presence of chemical substance (or a signal) $h(\bfm X_i, t)$ not only at the center $\bfm X_i$, but also in its proximity. To model this graded influence, a weighted average operator is defined:
\begin{equation}\label{eq:operator}
\mathcal{F}(h(\bfm X_i,t)) = \frac{1}{W}\int_{B(\bfm X_i,\bar{R})} w_i(\bfm x) h(\bfm x,t)\, d \bfm x,
\end{equation}
where:
\be
 W:=\int_{B(\bfm X_i,\bar R)} w_i(\mathbf{x})\,d\bfm x,  
\ee
and a possible choice for the weight function can be a truncated Gaussian 
\begin{equation*}
w_i(\bfm x) := 
\left\{\begin{array}{ll}
2 \exp\left(-|| \bfm x-\bfm X_i||^2 \displaystyle\frac{\log 2}{\bar R^2}\right) -1,  &\quad\textrm{if } || \bfm x- \bfm X_i|| \leq \bar R,\\
0,   &\quad\textrm{otherwise},
\end{array}
\right.
\end{equation*}
which models a sensing intensity decreasing on the distance from the cell center. Here $\bar R$ represents an influence radius, possibly greater than the cell radius $R$ (Table \ref{tab-param-dim}), and  
\begin{equation*}
B(\bfm X_i,R):=\{ \bfm x: || \bfm x- \bfm X_i || \leq R\} 
\end{equation*}
is the ball centered in $\bfm X_i$ with radius $R$. In the following the characteristic function over the ball $B(\bfm X_i,R)$ is defined as:
\begin{equation*}
\chi_{_{B(\bfm X_i,R)}} := \left\{\begin{array}{ll}
1,  &\quad \textrm{if }  \bfm x \in B(\bfm X_i,R), \\
0,   &\quad \textrm{otherwise}.
\end{array}\right.
\end{equation*}

\section{The mathematical model}\label{sec:model}

In this section we present in detail the discrete in continuous model for the formation of a Cardiosphere. Firstly, we present the equations describing the movement of cells and the forces acting on them. Then we focus on the proliferation and differentiation processes. The basic mechanisms of the phenomena are explained and the reaction-diffusion systems for the time evolution of the oxygen and TGF-$\beta$ concentrations are given.    

\subsection{Cells dynamics}

Motion of cells 
is the result of mechanical forces acting on them and of their reciprocal interactions. Due to the finite volume of cells there is an internal pressure between them which leads to repulsion 
while being compressed. On the other hand, there are attractive bonds between cells which keep them in contact. We also assume that cells are driven by a chemotactic signal, 
that is they move toward the higher gradient of the  TGF-$\beta$ concentration, 
and are slowed down because of the viscosity of the medium in which they are immersed \citep{rubinstein,fournier,bayly}. 

The position of the $i$-th cell $\bfm X_{i}$ follows  Newton's law and is described by the second order ordinary differential equation
\begin{equation}
\ddot {\bfm {X}}_{i} =\sum_{j:\mathbf{X}_j\in B(\mathbf{X}_i,R_2)\backslash\left\{\mathbf{X}_i\right\}}\bfm {K}( \bfm {r}_{ij}) - \mu \dot {\bfm {X}}_i + \alpha \mathcal{F}(\nabla S(\bfm{X}_i,t)) \label{eqn5},
\end{equation}
where dot denotes time derivative, $\mu$ is a friction coefficient for unit mass 
and $\bfm{ K}(\bfm {r}_{ij})$ is the attraction-repulsion force for unit mass between i-th and j-th cell, defined as
\begin{equation}\label{attr-rep}
{\bfm K(\bfm {r}_{ij})}:=\left\{
\begin{array}{ll}
\ds - k_1 \left(\ds\frac{1}{||\bfm r_{ij}||} - \ds\frac{1}{R_1}\right) \ds\frac{\bfm r_{ij}}{||\bfm r_{ij}||},&\quad ||\bfm r_{ij}||\leq R_1,\\
\\
\ds k_2(||\bfm r_{ij}||-R_1)\ds\frac{\bfm r_{ij}}{||\bfm r_{ij}||},&\quad R_1<||\bfm r_{ij}||\leq R_{2}.
\end{array}\right.
\end{equation}
The parameters $k_1$, $k_2$ are positive constants, $\bfm r_{ij}=\bfm X_j-\bfm X_i$ is the distance 
between cells, $R_1=2R$ and $R_2$ (to be chosen later, see Table \ref{tab-param-dim}) are the radii of repulsion and attraction respectively. Note that equation \eqref{attr-rep}$_1$ gives a repulsion in the form $1/r$, while \eqref{attr-rep}$_2$ is a linear 
attractive elastic force. 
Similar terms can be found in the model proposed in \citet{dorsogna,joie}.  
Equation (\ref{eqn5}) is augmented with the initial conditions
\begin{equation*}
\bfm X_i(0)=\bfm X_i^0,  \qquad \bfm{\dot X_i}(0)=\bfm V_i^0=\bfm {0}.
\end{equation*}

\subsection{Proliferation and differentiation}\label{sec:prol_diff}

Some of the present authors has been investigating the proliferation and differentiation dynamics of cells forming CSps from the original description of the method of isolation and expansion of human cardiac progenitors from heart biopsy \citep{mess}. Confocal analysis of the BrdU labeled cells which incorporate the base analogue within the replicating DNA, showed the positive fluorescence labeled cells located in the inner of the growing spheres. No or a weak signal was present in the external layers of more differentiated cells (that cannot replicate at the same speed). This kind of layered growth implies that the proliferation in the central area depends also on the CSps size and the core-cells may become quiescent or die by necrosis or apoptosis.

It is worth noting that tumor spheroid growth depends (at least partially) on the same factors, however, in this case they show a more pronounced variety of final volumes for any given cell line. We can distinguish  an early phase of global undifferentiated mass, a second stage of a shell of proliferating cells around a central area of quiescent elements and the last stage in which an inner necrotic/apoptotic core is enveloped by the previous two. The external layer always represents the growing front \citep{kim}. 

Stem cells are characterized by different levels of maturation (differentiation). To model this feature we mark the $i$-cell with a label $\varphi_i(t) = {1,2,3,d}$, called its ``state''. The state $\varphi=d$ indicates dead cells or the space occupied by necrotic mass (due to adverse biological conditions, as oxygen deficiency for a sufficiently long period of time).  Note that apoptosis, which is a programmed cells death, occurs over a longer timescale and is not considered here. In particular, we introduce a threshold concentration $c_d$ such that whenever $\mathcal{F}(c(\mathbf{X}_i,t))>c_d$ cell survives, otherwise it dies. Such cell is marked as dead ($\varphi=d$) and does not take part in further evolution of the CSp, that is it neither proliferates nor differentiates. However, dead cells do occupy a volume and exert attractive-repulsive forces on other cells. A stochastic process introduced in the process of proliferation and differentiation models also the presence of quiescent cells. Such cells neither proliferate nor differentiate but, in contrast to dead cells, this state results from environment conditions and is only temporary. At a later time a quiescent cell can restart its functions.  

For a living cell, $\varphi=1$ denotes the lowest degree of maturation (the least differentiated cells), while $\varphi=2,3$ are the intermediate and the highest degrees of differentiation, respectively. They can be experimentally reached with our previously cited method \citep{mess,chim1}. 
 Finally, we define a {\em cell cycle time $T_c$}, i.e. a cell maturation time after which the cell may proliferate and/or differentiate with some probability. 

\subsubsection{Mechanism of differentiation}\label{sec:diff}


In the model we assume that, with some probability and when {\em environment} conditions are satisfied, from time $t$ to time $t+\Delta t $ a cell can 
change its state to a higher one  
by a unity according to Table~\ref{Tab:differentiation}, or can die (as described in Section \ref{sec:proliferation}).

\begin{table}[htbp!]
\centering \begin{tabular}{|c||c|c|c|}
\hline
                & $\varphi_i(t+\Delta t)=1$ & $\varphi_i(t+\Delta t)=2$ & $\varphi(t+\Delta t)=3$ \\ \hline\hline
$\varphi_i(t)=1$& \checkmark                & \checkmark                &                         \\
\hline
$\varphi_i(t)=2$&                           & \checkmark                & \checkmark              \\
\hline
$\varphi_i(t)=3$&                           &                           & \checkmark
\\ \hline
\end{tabular}
\caption{Possible differentiation levels reached by a cell. The check marks the admissible transition at time $t+\Delta t$ for each cell state on the first column.}
\label{Tab:differentiation}
\end{table}
More precisely, the differentiation is triggered by chemical signals, such as TGF-$\beta$ present in the extracellular environment and secreted by cells themselves. Moreover, it can be inhibited due to the presence, in the surrounding area, of cells of the same type, as we can found in similar cell systems \citep{itoh,matsuda} and mathematical models (e.g. \citealp{dicostanzo1}). We denote by $\Gamma_i$ an inhibitor indicator function, which depends on the number $n_i$ of cells in the neighborhood of the i-th cell and is defined as
\begin{equation*}
\Gamma_i(t)=
\left\{
\begin{array}{ll}
1,  \quad & \text{if } n_i \leq \bar{n},  \\
0,  \quad & \text{if } n_i > \bar{n}, 
\end{array}
\right.
\qquad  n_i:=\text{card}\{j:\bfm X_j\in B(\bfm X_i,R_4)\},\qquad i=1,\dots,N.
\end{equation*}
For the threshold $\bar{n}$ we set in the following the value $\bar{n}=4$ for a two dimensional case. 

We define a differentiation threshold
\be \label{prob-q}
q_i(t):=
\displaystyle\frac{\sigma(\varphi_i) \cdot \mathcal{F} (S(\bfm X_i,t)) \cdot  \Gamma_i(t) }{ S_{\max}},  \qquad\qquad i=1,\dots,N,
\ee
for the cell $i$, with $\sigma(\varphi_i)$ ($\varphi_i=1,2$) positive constants, and $S_{\max}$ the experimental maximum value of TGF-$\beta$ (Table \ref{tab-param-dim}). In addition, we assume that differentiation occurs under the condition that oxygen concentration level is above a given threshold value $\bar c (\varphi_i)$, also depending on the cell states $\varphi_i=1,2$.  
\par The differentiation variable $\varphi_i(t)$ can be modeled as a non-homogeneous Poisson process with variable intensity $q_i(t)$ (see for example \citealp{anderson2015} and references therein, as well as \citealp{kushner}). In particular, the probability of a differentiation in a time interval $\Delta t$ is given by
\begin{align}
	P(\varphi_i(t+\Delta t)-\varphi_i(t)\geq 1)\approxeq q_i(t)\Delta t,
\end{align}
the expected number of differentiations in the interval $(t,t+\Delta t)$ is
\begin{align}
	E\left(\text{number of differentiations in $(t,t+\Delta t)$}\right)= q_i(t)\Delta t,
\end{align}
therefore in the time interval $(0,t)$ results
\begin{align}
	E\left(\text{number of differentiations in $(0,t)$}\right)= \int_0^t q_i(s)\,ds.
\end{align}
Finally, we write the stochastic equation
\begin{align}\label{eq-stoc-phi}
	d\varphi_i=d\mathcal{P}\left(\int_0^t q_i(s)\,ds\right),\quad i=1,\dots,N,
\end{align}
$\mathcal{P}(\int_0^t q_i(s)\,ds)$ being a inhomogeneous Poisson process with intensity $q_i(t)$.
\par In practice,  if $\mathcal{F} (c(\bfm X_i,t))\geq \bar c (\varphi_i)$, starting from \eqref{eq-stoc-phi}, we compute the threshold \eqref{prob-q}, and we can adopt the following stochastic rule: cell types $\varphi_i=1$ and $\varphi_i=2$ switch irreversibly their state according to 
\begin{equation} 
\varphi_i(t+\Delta t) = \varphi_i(t)+
\left\{
\begin{array}{ll}
1, &\quad\textrm{if } \bar q_i < q_i \Delta t, \\
0, &\quad\textrm{otherwise},
\end{array}\right.    \qquad i=1,\dots,N,
\end{equation}
where $\bar{q}_i$ is a random number between 0 and 1 (see also Section \ref{sec:numeric}). 

\subsubsection{Mechanism of proliferation}\label{sec:proliferation}

\begin{figure}[htbp!]
\centering\includegraphics[scale=.70, angle=0]{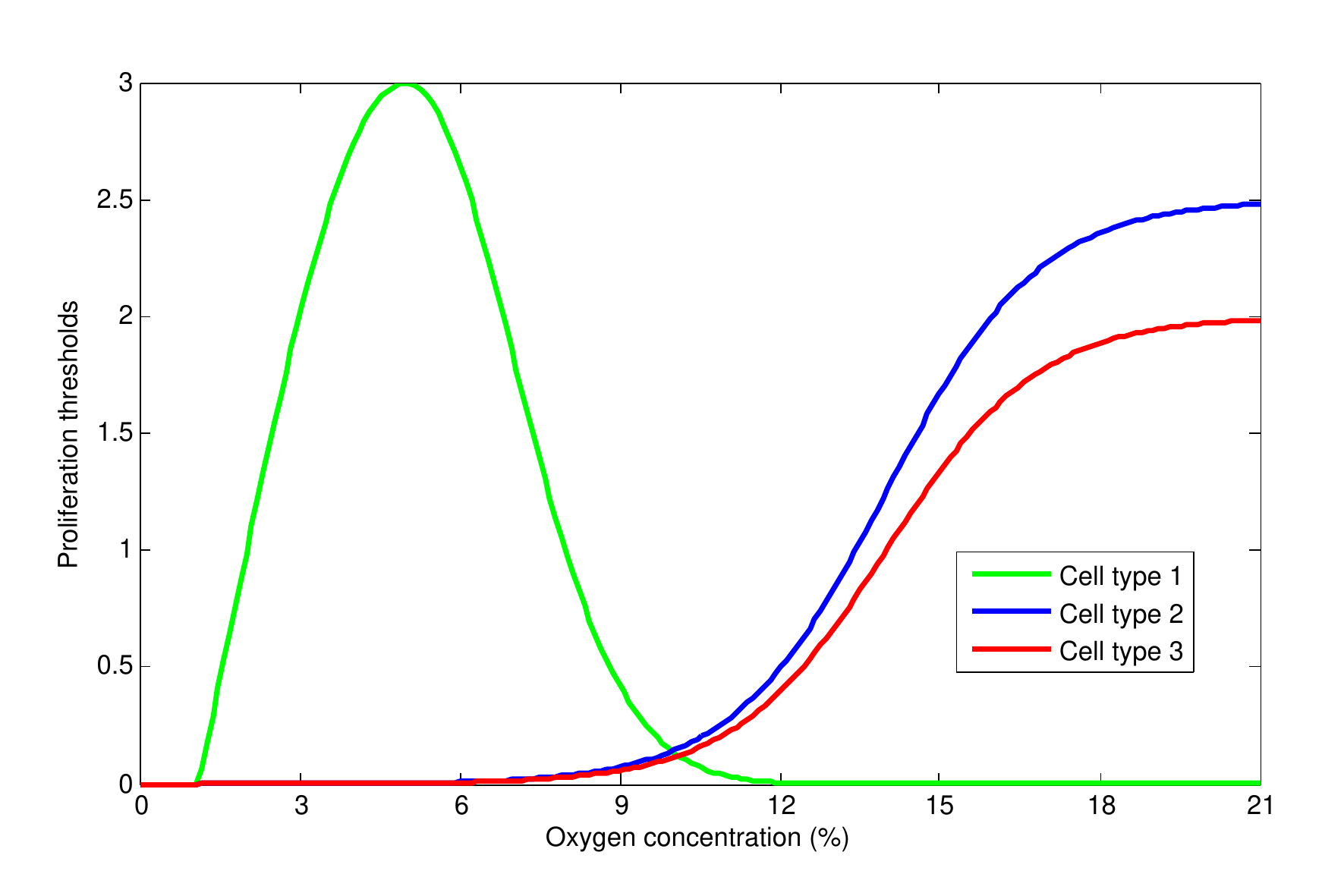}  
\caption{Proliferation thresholds as a function of oxygen concentration (equation \eqref{prob}).} 
\label{fig4}    
\end{figure}

We introduce the proliferation thresholds $p_i(t)$ of the i-th cell at time $t$ in the following form 
\begin{equation}\label{prob}
p_i(t) = \left\{\begin{array}{ll}
g_1(\mathcal{F}(c)),&\quad\textrm{if }\varphi_i(t)=1,\\
g_2(\mathcal{F}(c)),&\quad\textrm{if }\varphi_i(t)=2,\\
g_3(\mathcal{F}(c)),&\quad\textrm{if }\varphi_i(t)=3.\\
\end{array}\right.
\end{equation}
Functions $g_1, g_2, g_3$  describe the dependence of the rate of proliferation on the concentration of oxygen. Least differentiated cells, i.e. those at state $\varphi=1$, work under anaerobic conditions. Their proliferative ability increases at low levels of oxygen with an assumed gaussian-type dependence. When $c$ is too high, cells change their metabolism and proliferate at a lower rate. Cells at states $\varphi=2,3$ behave on the opposite: their proliferation rate increases with higher concentration of oxygen. However, due to the fact that they are already differentiated, their ability to proliferate is lower than the stem cells at $\varphi=1$. As a consequence, a possible choice for functions $g_1, g_2, g_3$, used in the numerical simulations, is shown in Figure~\ref{fig4}. If, for certain interval of time, the concentration of oxygen is below a given threshold $c_d$, then cells do not have enough resources to live and die. \par
To model stochasticity in proliferation, we introduce a new state variable $\phi_i(t)$ denoting the number of 
(from $i-th$) generated cells  in $(0,t)$. 
Similarly to the differentiation (Section \ref{sec:diff}),  $\phi_i(t)$ satisfies a non-homogeneous Poisson process: 
\be
d\phi_i=d\mathcal{P}\left(\int_0^t p_i(s)\,ds\right),\quad i=1,\dots,N,  \label{ery}
\ee
with variable intensity $p_i(t)$ given by equation (\ref{prob}).

If $\mathcal{F}(c(\mathbf{X}_i))>c_d$, we compute the proliferation threshold of the i-th cell $p_i$ through equation (\ref{prob}). The following computational rule of proliferation is applied:

\begin{equation} 
\phi_i(t+\Delta t) = \phi_i(t)+
\left\{
\begin{array}{ll}
1, &\quad\textrm{if } \bar p_i < p_i \Delta t, \\
0, &\quad\textrm{otherwise},
\end{array}\right.    \qquad i=1,\dots,N,
\end{equation}
where $\bar{p}_i$ is a random number between 0 and 1 (see also Section \ref{sec:numeric}). 

%
 When a cell proliferates, a newborn $(N+1)$-th cell appears and is appended to the vector of coordinates $\bfm X_i$, $i=1,2,\dots,N$. Its position $\bfm X_{N+1}$ is generated randomly within the distance $r$ from the mother cell 
 and with same velocity. More precisely, in polar coordinates we choose randomly two numbers: the angle $\theta\in[0,2\pi]$ and the distance between the two centers $r\in[R/2,R]$. 

\subsection{Dynamics of oxygen and TGF-\texorpdfstring{$\beta$}{beta} signal}

The concentration of oxygen $c=c(\mathbf{x},t)$ is a continuous variable and its time evolution is governed by a reaction-diffusion equation:
\begin{equation}\label{eqnoxy}
\left\{
\begin{array}{ll}
\ds \partial_{t} c = \nabla \cdot (D_c \nabla c) -  \sum_{i=1}^{N}\frac{ \lambda (\varphi_i) c^{\gamma+1}}{k(\varphi_i) + c^{\gamma+1} } \chi_{_{B(\mathbf{X}_i,R_3)}} + H \bar{B} (c_0 -c),&   \\\\
\ds c(\mathbf{x},0)=c_0,  &\quad \text{in }  \Omega, \\\\
\mbox{periodic boundary conditions,} &\quad \text{on }  \partial\Omega,
\end{array}\right.  
\end{equation}
with $c_0$ the environmental (constant) concentration. The last term in equation (\ref{eqnoxy})$_1$  
models the two-dimensional representation of the real 3D experiment: the 
supply of oxygen coming from the environment (upper surface of the culture) is recovered as a source term, modeled as proportional to the difference $c_0-c$ with rate $H$,  
and spatially modulated by a well-shaped function over the CSp geometry, in the form 
\begin{align}\label{eq:well}
\bar{B} (\bfm x):=  
\left\{
\begin{array}{ll}
\displaystyle \frac{\exp \left[\zeta \left( \ds \frac{r(\bfm x)}{ R_{\text{CSp}}}
\right)^2 \right] -1 }{\exp(\zeta) -1},&\quad \text{if }r\leq R_{\text{CSp}},\\
1,&\quad \text{otherwise}.
\end{array}
\right.
\end{align}
The constant $\zeta$ is a constant geometrical parameter, $r(\bfm x)$ is the distance from the center of the sphere, and $R_\text{CSp}$ is the mean radius of CSp at time $t$. The shaped function \eqref{eq:well} considers the graded exposition to oxygen of cells at different depth. In the equation (\ref{eqnoxy}) we 
assume that oxygen diffuses in the environment  and is consumed by cells with a rate depending on the metabolism of the cell (i.e. on its state $\varphi_i$), and 
increases with the oxygen availability ($ \propto c^{\gamma}, \, \gamma>0$) according to the saturation Michaelis-Menten-like 
law \citep{wu}. A typical value $\gamma=1/2$ is assumed.  Degradation of the oxygen is negligible over the typical time scale (few days) of the experiment.

The concentration of the chemical signal $S(\mathbf{x},t)$ is governed by the similar IBV problem
\begin{equation}\label{eqntgf}
\left\{
\begin{array}{ll}
\partial_{t} S = \nabla \cdot ( D_{S} \nabla S) + \ds\sum_{i=1}^{N} {\xi (\varphi_i) \chi_{_{B(\mathbf{X}_i,R_3)}}} - \eta S,  & \\\\  
S(\mathbf{x},0)=S_0,   &\quad \text{in }  \Omega,       \\\\ 
\mbox{periodic boundary conditions,} &\quad \text{on }  \partial\Omega,
\end{array} 
\right.
\end{equation}
where $\xi$  is  the (state dependent) $S$-release  rate, $\eta$ is the molecular degradation rate and $D_S(\bfm x)$ is a diffusion coefficient.
 Compared with the production term, a possible $S$-consumption term is extremely low and has been neglected. 

In addition,  the diffusivity coefficient $D$ (say $D_c$ in equation (\ref{eqnoxy}),  $D_S$ in (\ref{eqntgf})) varies in relation with the change of cell density, which we measured 
in a reference volume. 
Due to the presence of a large number of cells, $D$ is significantly reduced at the center of the CSp. Diffusion can be considered as in a porous medium, and the \emph{effective diffusivity}
$D(\bfm x)$ can be defined as:  \\
\bdm
D(\bfm x):= \frac{D^{\max}}{1 + \rho \bar A(\bfm x)},
\edm
where $D^{\max}$ is the constant unperturbed diffusion, $\rho$ is a compactness parameter and $\bar A (\bfm x)$ is the occupancy fraction in a proper control volume containing $\bfm x$. 

\section{Numerical results and discussion}\label{sec:numeric}

In this section we present numerical simulations of a typical formation and growth of CSps, under typical conditions \citep{mess,chim1}, using the previously described mathematical model. 

For convenience we summarize here our hybrid system of equations:
\begin{align}\label{eq:complete}
	\left\{
\begin{array}{ll}
\ddot {\mathbf{X}}_i &=\displaystyle\sum_{j:\mathbf{X}_j\in B(\mathbf{X}_i,R_2)\backslash\left\{\mathbf{X}_i\right\}}\bfm {K}( \bfm {r}_{ij}) - \mu \dot {\bfm {X}}_i + \alpha \mathcal{F}(\nabla S(\bfm{X}_i)), \\\\
d\varphi_i&=d\mathcal{P}\left(\displaystyle\int_0^t q_i(s)\,ds\right), \\\\
d\phi_i&=d\mathcal{P}\left(\displaystyle\int_0^t p_i(s)\,ds\right),  \\\\
\ds \partial_{t} c &=\displaystyle \nabla \cdot (D_c \nabla c) - \displaystyle \sum_{i=1}^{N}\frac{ \lambda (\varphi_i) c^{\gamma+1}}{k(\varphi_i) + c^{\gamma+1} } \chi_{_{B(\mathbf{X}_i,R_3)}} + H \bar{B} (c_0 -c),  \\\\
\partial_{t} S &=\displaystyle \nabla \cdot ( D_{S} \nabla S) + \ds\sum_{i=1}^{N} {\xi (\varphi_i) \chi_{_{B(\mathbf{X}_i,R_3)}}} - \eta S,   
\end{array}
\right.
\end{align}
where $\bfm {K}( \bfm {r}_{ij})$ and $\mathcal{F}(\nabla S(\bfm{X}_i))$ are given respectively in \eqref{attr-rep} and \eqref{eq:operator}, while $q_i$ and $p_i$ given by \eqref{prob-q} and \eqref{prob}. Initial and boundary conditions are given by:
\begin{align*}\label{boundary1}
	\bfm X_i(0)&=\bfm X_i^0,  \qquad \dot{\bfm{ X}}_i(0)=\bfm {0},\\
	N(0)&=N_0,\qquad\varphi_i(0)=1, \qquad \forall i=1,\dots,N_0,\\
	c(\mathbf{x},0)&=c_0,   \qquad \text{periodic boundary conditions on $\partial\Omega$},\\
	S(\mathbf{x},0)&=S_0,   \qquad \text{periodic boundary conditions on $\partial\Omega$}.
\end{align*}

Initially, we consider a cluster of $N_0=15$ cells at the first stage of differentiation ($\varphi_i=1$, $\forall i$) placed at the center of a square domain $\Omega=[0,225]\times [0,225]\,(\mu\text{m}^2)$. Such setting models the initiation of the numerical simulation at the time that corresponds to about 24 hours of the real experiment. 

Despite the low geometrical dimension and a number of simplifying assumptions, the large set of biological parameters influences the problem and their complete characterization remains a difficult task. Each parameter is interconnected with the others and strongly affects the simulation. Some of them are taken from previously published results or literature, others are chosen in a compatible and consistent range, and the remaining ones are varied to analyze the sensitivity of the system. The complete set of biological parameters is given in Table~\ref{tab-param-dim}.

\subsection{Numerical methods}\label{numerical_method}

All the equations are discretized by finite differences. The second order equation (\ref{eq:complete})$_1$ is reduced to a system of first order equations and solved by
explicit Euler in time. In equations (\ref{eq:complete})$_{4,5}$ the diffusion terms are discretized by a standard centered difference in space and integrated implicitly in time. 
The nonlinear reaction terms are treated explicitly in time. Finally, in equation (\ref{eq:complete})$_{5}$ we have eliminated the stiff term  $-\eta S$ using a classical exponential transformation.

The square domain size is chosen sufficiently large, $225\, \mu$m per side, so that the growing CSp does not reach the boundary in the typical time of observation. We consider an uniform mesh with grid spacing $ \Delta x= \Delta y=3.75\, \mu$m, which guarantees a sufficient discretization over each cell (having a diameter of 15 $\mu$m, see Table~\ref{tab-param-dim}) and a reasonable accuracy.  

The time step $\Delta t$ has been fixed as the maximum value to ensure stability and non-negativity of the scheme ($\max \Delta t= 0.02$ h). For the same reason, when the oxygen 
concentration drops below a given threshold, it is necessary to reduce the time step to $\Delta t=0.001$ h. 

\subsection{Parameter sensitivity analysis}\label{sec:sensitivity}

Mathematical models of biological phenomena are subject to sources of uncertainty, including errors of measurement, natural intrinsic variability of the system, absence of information and poor or partial understanding of the driving forces and mechanisms. This uncertainty imposes a limit on our confidence in the response or output of the model. One of the challenges is to characterize the model parameters that are not identified experimentally, and these parameters are marked as \emph{calibrated} in Table \ref{tab-param-dim}. 
\par In this section we present the calibration procedure in order to obtain as reliable results as possible.  To achieve this goal, we studied the influence of the parameters on the model dynamics through a local sensitivity analysis \citep{saltelli,clarelli}. This approach determines a degree of dependency between input parameters (one at a time) and the results of simulations. Although our analysis does not take into account the presence of interactions between parameters, as in the global sensitivity analysis, it can give useful information for further exploration of the parameter space. 

We measure the sensitivity of the model to the variation of a positive quantity $\Psi$ with respect to a reference parameter $p_0$ by defining a sensitivity index  $SV \geq0$ as:
\be\label{eq:SV}
SV:=\frac{ |\bar \Psi(p_0 \pm \varepsilon)- \bar \Psi(p_0)| / \bar \Psi(p_0)}{\varepsilon /p_0},
\ee
where $\bar \Psi$ is the average value of $\Psi$ over a large number 
of runs (100 in our analysis) and it accounts for the stochasticity presents in 
our model. Typically, $\varepsilon$ is a small deviation over $p_0$ ($\varepsilon=0.05 p_0$, i.e. $5\%$ variation). In Table~\ref{tab:sensitivity_index} we report the sensitivity index computed with respect to two observed variables: the total number of cells $N$, and the diameter of the CSp, both at the final simulated time of 72 hours. The smallness of the index give us confidence of the robustness of the model in the response of the system in the presence of uncertainty.
\begin{landscape}
\begin{longtable}{@{\extracolsep{\fill}}*{4}{l}}
\caption{Estimates of physical and biological parameter values. Whenever possible, the values are taken from literature or experiments, the remaining are calibrated to be consistent with other (c.w.o.). Their influence on the system is studied through a local sensitivity analysis (Section \ref{sec:sensitivity}). When a parameter admits a range of possible values, the used one in the numerical tests is put in brackets.}
\label{tab-param-dim}\\
\toprule
\multicolumn{1}{l}{Parameter} & \multicolumn{1}{l}{Definition} & 
\multicolumn{1}{l}{Estimated value or range (used values)} & \multicolumn{1}{l}{Source} \\
\midrule
\endfirsthead

\multicolumn{2}{l}{\footnotesize\itshape\tablename~\thetable:
continuation from the previous page} \\
\toprule
Parameter & Definition                                                & Estimated value or range (used values)   & Source\\
\midrule
\endhead

\midrule
\multicolumn{2}{r}{\footnotesize\itshape\tablename~\thetable:
continuation in the next page} \\
\endfoot

\bottomrule
\endlastfoot

$R$       & \scriptsize cell radius                                      & 7.5 $\mu\text{m}$                        & \scriptsize\citet{wu}\\

$\bar{R}$ & \scriptsize detection radius of chemicals                    & 2R                                       &\scriptsize biological assumption \\
$R_1$     & \scriptsize radius of action of repulsion between cells      & 2R                                       &\scriptsize biological assumption\\
$R_2$     & \scriptsize radius of action of adhesion between cells       & 2.5R                                     &\scriptsize biological assumption \\
$R_3$     & \scriptsize radius of production/degradation of chemicals    & R                                        &\scriptsize biological assumption\\
$R_4$     & \scriptsize detection radius for the differentiation inhibition  &2R                                        &\scriptsize biological assumption \\
$\alpha$  & \scriptsize coefficient of chemotactic effect per unit mass  & $10^{10} \;\mu\text{m}^{4}\,\text{h}^{-2} \text{pg}^{-1}$                                                         &\scriptsize calibrated, c.w.o.\\
$k_{1}$   & \scriptsize coefficient of repulsion per unit mass           & $10^{17}\;\mu\text{m}^{2}\,\text{h}^{-2}$                                                                         &\scriptsize calibrated, c.w.o.\\
$k_{2}$   &\scriptsize elastic constant per unit mass                    & \parbox[t][][t]{7cm}{$1.29 \times 10^{14}\text{ -- }1.29 \times 10^{19}$ $(1.29\times 10^{14})\;\text{h}^{-2}$} & \scriptsize \citet{bell} \\
$\mu$    &\scriptsize friction coefficient per unit mass                 & $5.82\times 10^{14}\text{ -- }(5.82\times 10^{15}) \; \text{h}^{-1}$                                               & \scriptsize \citet{rubinstein} \\
$D_c^{\max}$ & \scriptsize oxygen diffusion coefficient                   & $3.72\text{ -- } 4.93 \, (4.32) \times 10^6 \mu \text{m}^2\,\text{h}^{-1}$                                        & \scriptsize \citet{wu}\\
$D_S^{\max}$ & \scriptsize TGF-$\beta$ diffusion coefficient              & $9.36 \times 10^4  \mu \text{m}^2\,\text{s}^{-1}$                                                                 &\scriptsize \citet{sito} \\
$\xi (\varphi_{1,2,3})$ & \scriptsize coefficient of production of TGF-$\beta$ & $2.7 \times10^{-8} \text{ -- } 1.1 \times 10^{-6}$ $(5.64 \times 10^{-7}) \;\text{pg}\,\mu \text{m}^{-2}\,\text{h}^{-1}$ & \scriptsize \citet{chim}\\
$S_{\max}$ & \scriptsize maximum concentration of TGF-$\beta$                                 & $3 \times 10^{-8} \; \text{pg} \, \mu \text{m}^{-2}$                                                               &\scriptsize \citet{chim}\\
$\eta$   & \scriptsize degradation constant of TGF-$\beta$              & $13.86\text{ -- }20.79$ $(17.33)\;\text{h}^{-1}$\;                                                                    &\scriptsize \citet{wake}\\
$\sigma(\varphi=1,2)$   & \scriptsize differentiation constant         & 50 (if $\varphi=1$), 2.5 (if $\varphi=2$) (nondim.)                                                                                                          & \scriptsize calibrated, c.w.o.\\
$\lambda(\varphi=1,2,3)$ & \scriptsize oxygen consumption constant     &\parbox[t][][t]{6.5cm}{ $ 0.9 \times 10^{-2} \text{ -- } 2.52 \times 10^{-2}$  \\(1.4 $\times 10^{-2}$ if $\varphi=1$, 2.5$\times 10^{-2}$ if $\varphi=2$, 2.5 $\times 10^{-2}$ if $\varphi=3$) $\text{pg}\, \mu \text{m}^{-2} \, \text{h}^{-1}$ }& \scriptsize \citet{wu} \\
$k(\varphi=1,2,3)$       &\scriptsize Michaelis-Menten oxygen constant & $1.67 \times 10^{-5} \; \text{pg}\, \mu \text{m}^{-2}$                                                              &\scriptsize \citet{wu} \\

$\bar c (\varphi=1,2)$  &\scriptsize oxygen concentration threshold value  for diff.  & 4$\times 10^{-4}$ (if $\varphi=1$), 28$\times 10^{-4}$ (if $\varphi=2$)  $\text{pg}\, \mu \text{m}^{-2}$                                              &\scriptsize calibrated, c.w.o.\\
$c_{d}$  &\scriptsize oxygen minimum concentration for life (necrosis)  & $1.93  \times 10^{-4}\;  \text{pg}\, \mu \text{m}^{-2}$                                                                &\scriptsize \citet{wu} \\

\newpage

$c_0$    & \scriptsize environmental concentration  for oxygen          & \parbox[t][][t]{6cm}{$3.68   \times 10^{-3}\, \text{pg}\, \mu \text{m}^{-2}  \mbox{ (if $O_2$ at 21\%)}$, $8.83   \times 10^{-4}\, \text{pg}\, \mu \text{m}^{-2}  \mbox{ (if $O_2$ at 5\%)}$ } &\scriptsize \citet{wu} \\
$S_0 $   & \scriptsize  environmental concentration for S              & $6.62 \times 10^{-9}   \, \text{pg}\, \mu \text{m}^{-2}$                                                             &\scriptsize \citet{ther} \\
$T_c $   & \scriptsize cell cycle time                                 & $15\,\text{h}\mbox{ (if $O_2$ at 21\%)},$ $12\,\text{h} \mbox{ (if $O_2$ at 5\%)}$                                        &\scriptsize  \citet{mess}  \\
$ H $    & \scriptsize source rate                                     & $40\,\text{h}^{-1}$                                                                                                         &\scriptsize calibrated, c.w.o.  \\
$\rho $  &\scriptsize compactness parameter                            & $9\times 10^{3}$ (nondim.)                                                                                                   &\scriptsize  calibrated, c.w.o. \\
\end{longtable}
\end{landscape}
\begin{table}[htbp!]
\caption{Local sensitivity index $SV$ as defined in equation \eqref{eq:SV} with $\varepsilon$ corresponding to a 5\% variation, computed for the parameters used in the numerical simulations and marked as calibrated in Table \ref{tab-param-dim}. The average values have been taken on 100 independent runs, in the case of oxygen at 21\%. Observed variables $N$ and the CSp diameter have been considered at the final time of 72 h. In the first row for $N$ and for the diameter the reference average value and the standard deviation are shown.}
\label{tab:sensitivity_index}
\bc
\begin{tabular}{l l l l l}
\toprule
 Parameters changed               & $N$           & $SV_N$  & diameter     & $SV_{\text{diameter}}$  \\
                                  &  {\footnotesize average=73.31} &  & {\footnotesize average=123.28 $\mu$m} & \\
																	 &  {\footnotesize (st. dev.=13.42 \%)} &  & {\footnotesize (st. dev.=5.71\%)} & \\
\midrule
 $\alpha + \varepsilon $             &  -0.68 \%     &  0.14   &  -0.21\%   &   0.04  \\
$\alpha - \varepsilon $              &  -1.28 \%     &  0.26   &  -0.83\%   &   0.17  \\
$\bar{c}(\varphi_1) + \varepsilon $  &  -2.08 \%     &  0.41   &  -1.38\%   &   0.27  \\
$\bar{c}(\varphi_1) - \varepsilon $  &  -1.28 \%     &  0.26   &  -0.83\%   &   0.16  \\
$\bar{c}(\varphi_2) + \varepsilon $  &  +0.31 \%     &  0.06   &  +0.36\%   &   0.07  \\
$\bar{c}(\varphi_2) - \varepsilon $  &  +0.29 \%     &  0.06   &  -0.51\%   &   0.10  \\
$H + \varepsilon $                   &  -0.29 \%     &  0.06   &  -0.30\%   &   0.06  \\
$H - \varepsilon $                   &  +2.74 \%     &  0.55   &  -0.73\%   &   0.14  \\
$k_1 + \varepsilon $                 &  +0.79 \%     &  0.16   &  +0.40\%   &   0.08  \\
$k_1 - \varepsilon $                 &  -0.28 \%     &  0.05   &  -2.57\%   &   0.51  \\
$k + \varepsilon $                   &  +0.87 \%     &  0.17   &  -0.05\%   &   0.01  \\
$k - \varepsilon $                   &  -0.25 \%     &  0.05   &  -1.19\%   &   0.24  \\
$\sigma(\varphi_1) + \varepsilon $   &  -1.36 \%     &  0.27   &  -1.15\%   &   0.23  \\
$\sigma(\varphi_1) - \varepsilon $   &  -1.87 \%     &  0.37   &  -1.43\%   &   0.29  \\
$\sigma(\varphi_2) + \varepsilon $   &  -2.63 \%     &  0.53   &  -0.83\%   &   0.17  \\
$\sigma(\varphi_2) - \varepsilon $   &  +0.87 \%     &  0.17   &  -0.83\%   &   0.17  \\
$\rho+ \varepsilon $                 &  +0.80 \%     &  0.16   &  -0.70\%   &   0.13  \\
$\rho - \varepsilon $                &  -0.53 \%     &  0.10   &  -0.76\%   &   0.15  \\
$\lambda(\varphi_1)+ \varepsilon $   &  +1.32 \%     &  0.26   &  -1.55\%   &   0.31  \\
$\lambda(\varphi_1) - \varepsilon $  &  -2.55 \%     &  0.51   &  -0.84 \%  &   0.16  \\
$\lambda(\varphi_2)+ \varepsilon $   &  +1.80 \%     &  0.36   &  -0.15\%   &   0.03  \\
$\lambda(\varphi_2) - \varepsilon $  &  -2.01 \%     &  0.40   &  -1.01 \%  &   0.20  
\\
\midrule
\end{tabular}
\ec
\end{table}

\subsection{Simulations: growth and maturation of the CSp}
We performed numerical simulations of growth and differentiation of CSp considering  the $O_2$ and TGF-$\beta$ as the only key-regulatory biological mechanism. We compared the composition and the structure of the CSp at two typical experimental oxygen conditions. 

\begin{figure}[htbp!]
\centering
\subfigure[]{\includegraphics[scale=0.5, angle=0]{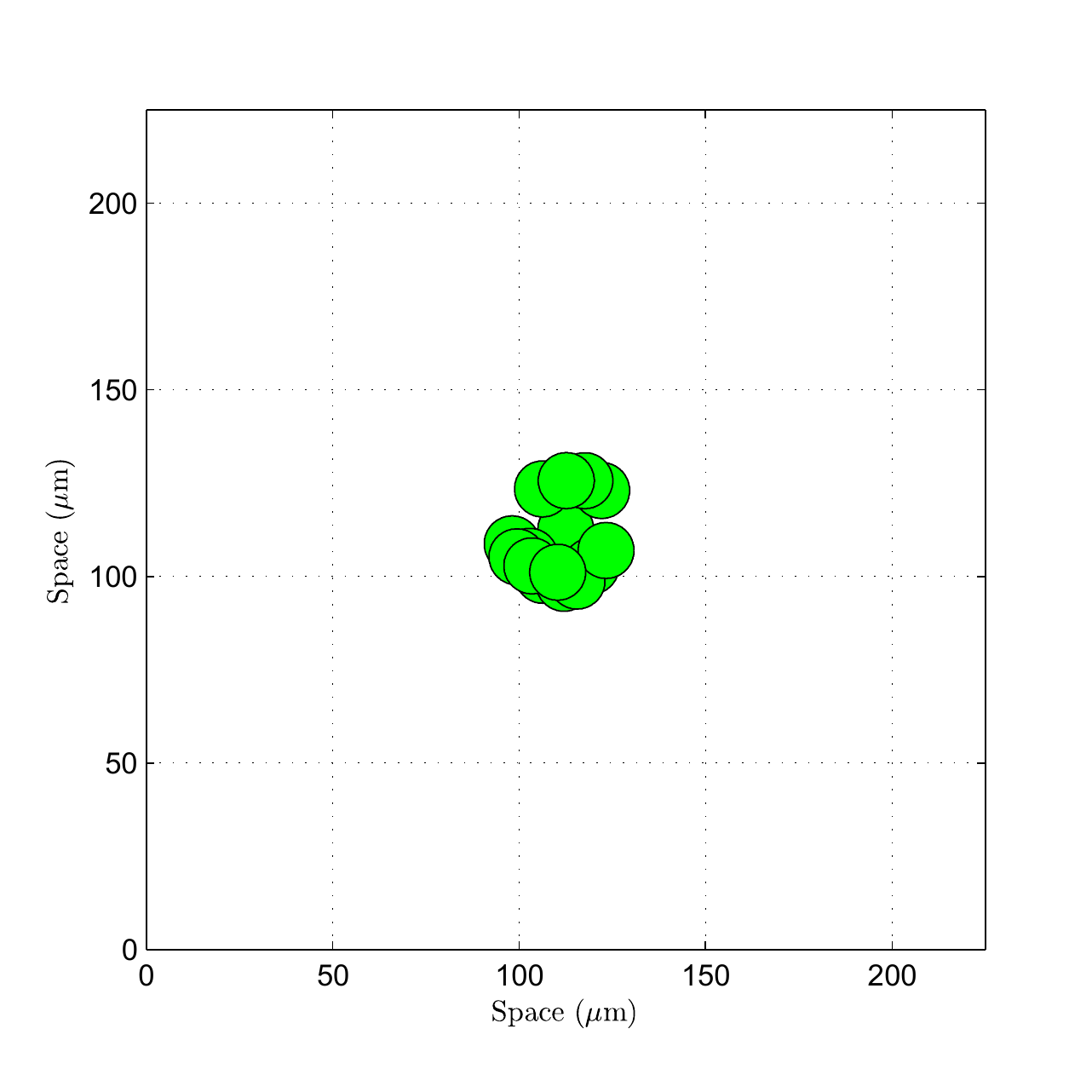}}  
\subfigure[]{\includegraphics[scale=0.5, angle=0]{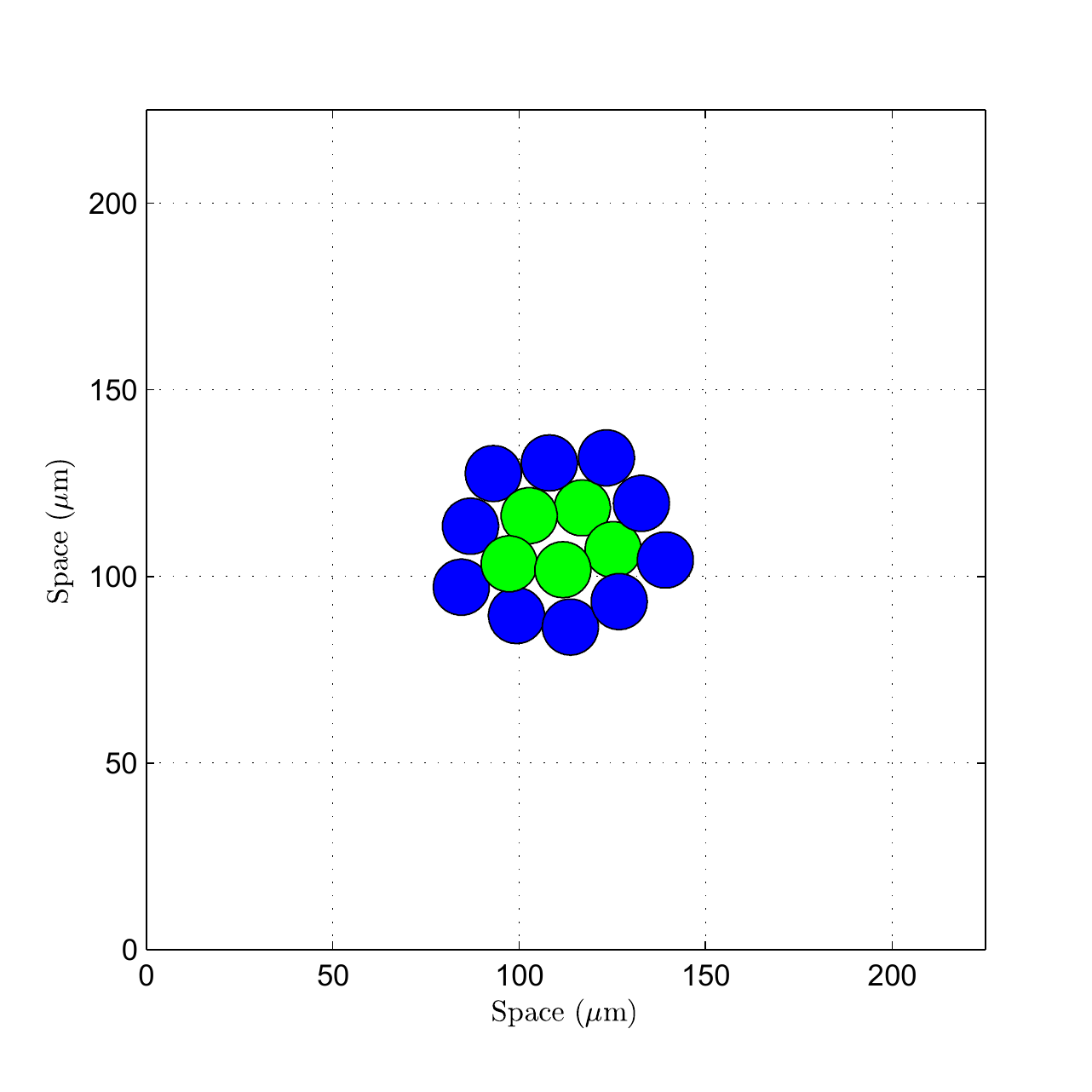}}  
{\scriptsize $t=0,  N_1=15, N_2=0,  N_3=0,  N_d=0$}  \qquad \qquad  
{\scriptsize $t= 24h,  N_1=5, N_2=10,  N_3=0,  N_d=0$} 
\subfigure[]{\includegraphics[scale=0.5, angle=0]{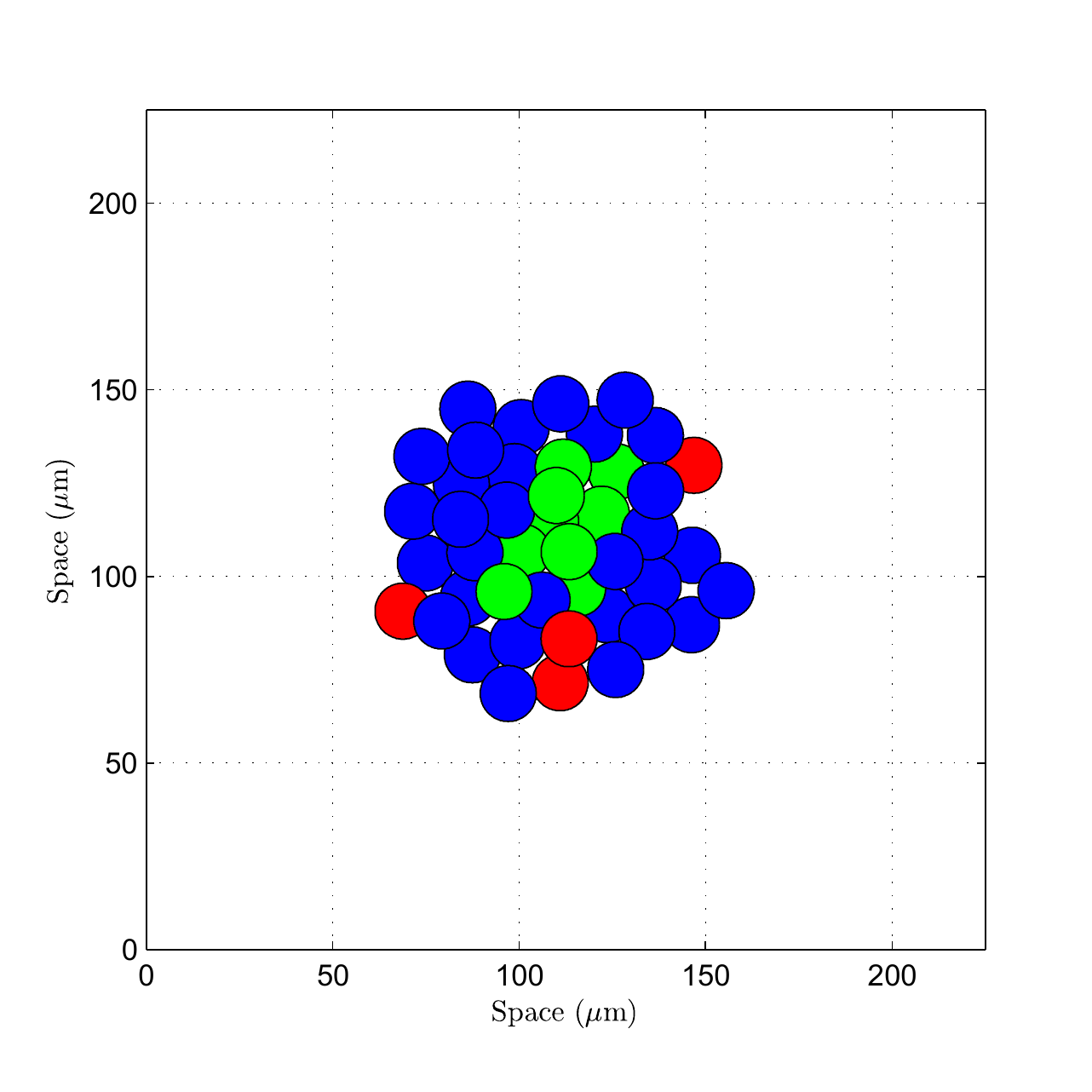}}  
\subfigure[]{\includegraphics[scale=0.5, angle=0]{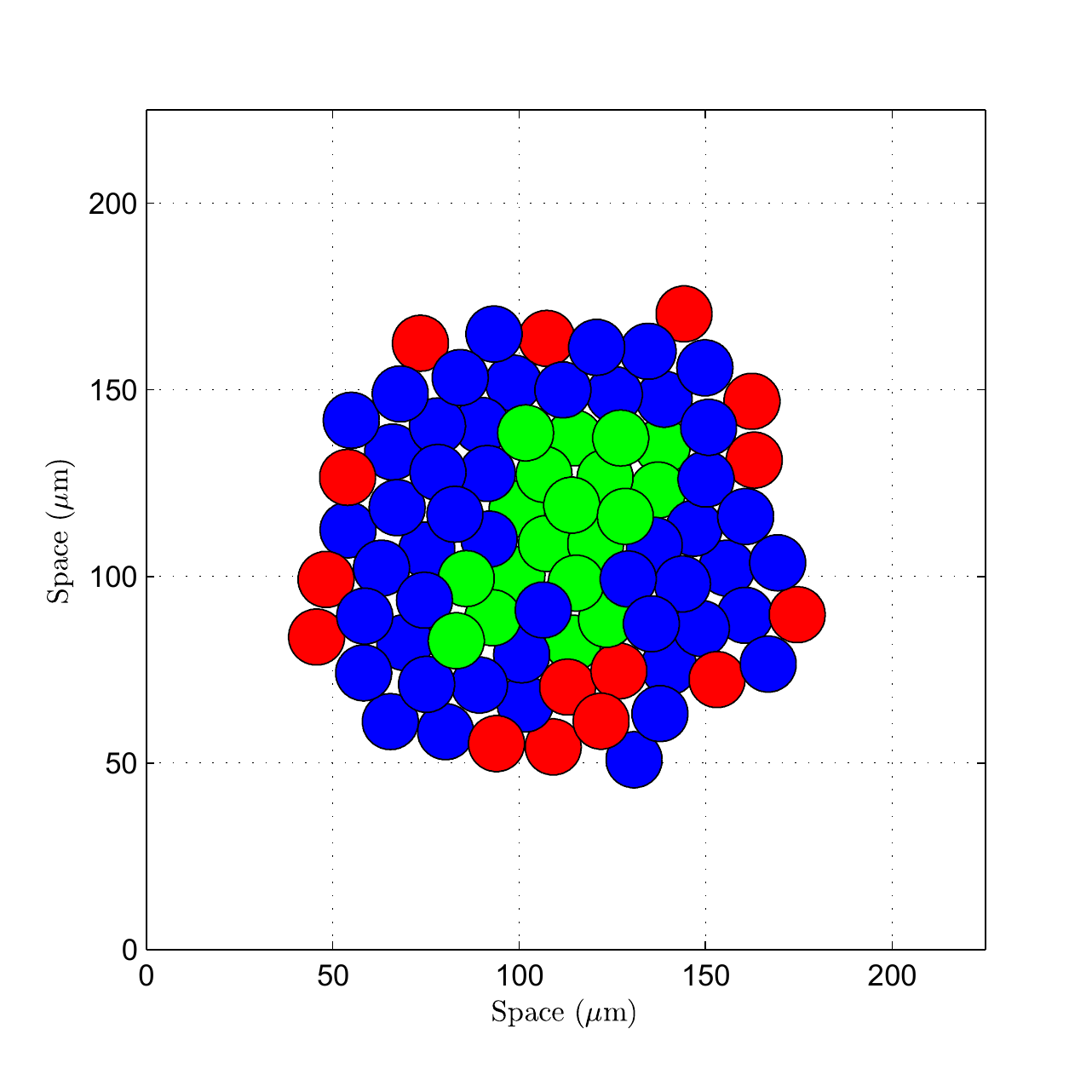}} \\
{\scriptsize $t=48h,  N_1=9, N_2=31,  N_3=4,  N_d=0$}  \qquad  \qquad  
{\scriptsize $t=72h, N_1=19, N_2=49,  N_3=15,  N_d=0$}
\caption{(a)--(d) Numerical simulation of the CSp growth for a $21\%$ oxygen environmental concentration, at times $t=0, 24, 48, 72$ h. System \eqref{eq:complete} is solved in a domain $\Omega=\left[0,225\right]\times\left[0,225\right]$ ($\mu$m$^2$), 
with the model parameters given by Table~\ref{tab-param-dim}. 
Differentiation levels are marked by different colors: green for the least differentiated cells ($\varphi=1$), blue color is for the intermediate level of differentiation ($\varphi=2$), while the red color labels cells with the highest degree of maturation ($\varphi=3$). Values $N_1$, $N_2$, $N_3$, $N_d$, in the subplot captions indicate the numbers of cells in the CSp, respectively for each state of maturation and for dead cells, $\varphi=1,2,3,d$.} 
\label{fig:sphere_21percent}  
\end{figure}

\begin{figure}[htbp!]
\centering
\subfigure[]{\includegraphics[scale=0.35, angle=0]{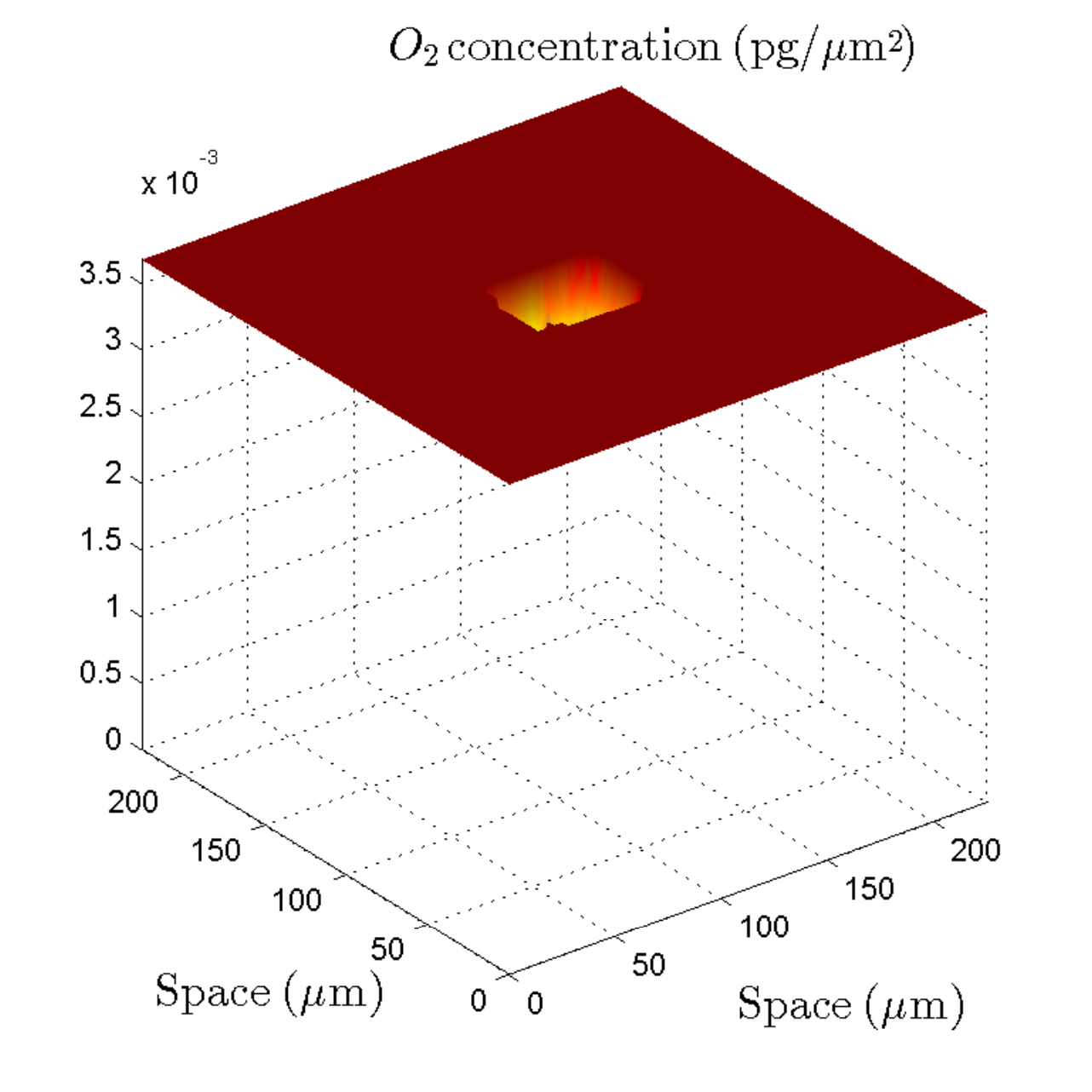}}  
\subfigure[]{\includegraphics[scale=0.35, angle=0]{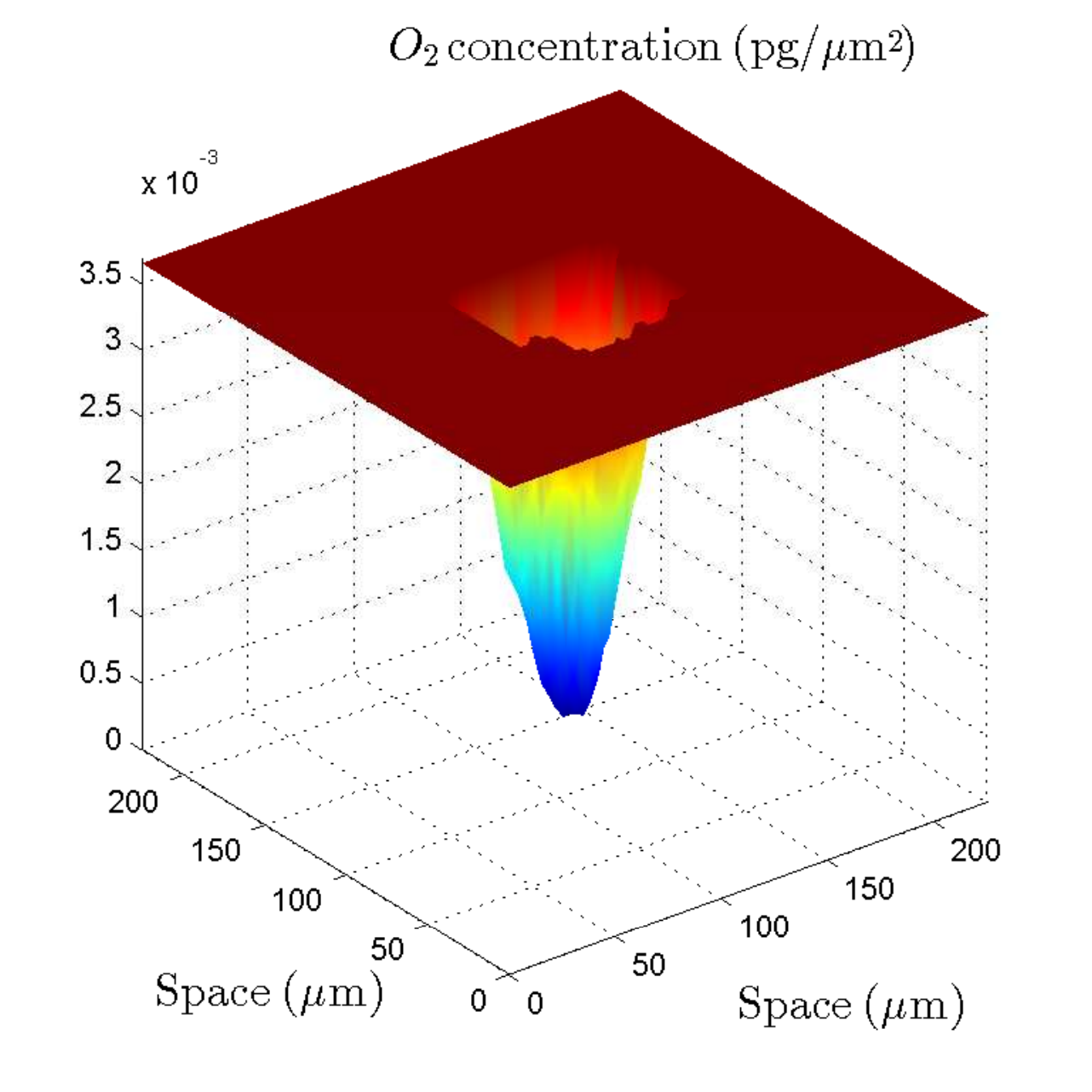}}  
\subfigure[]{\includegraphics[scale=0.35, angle=0]{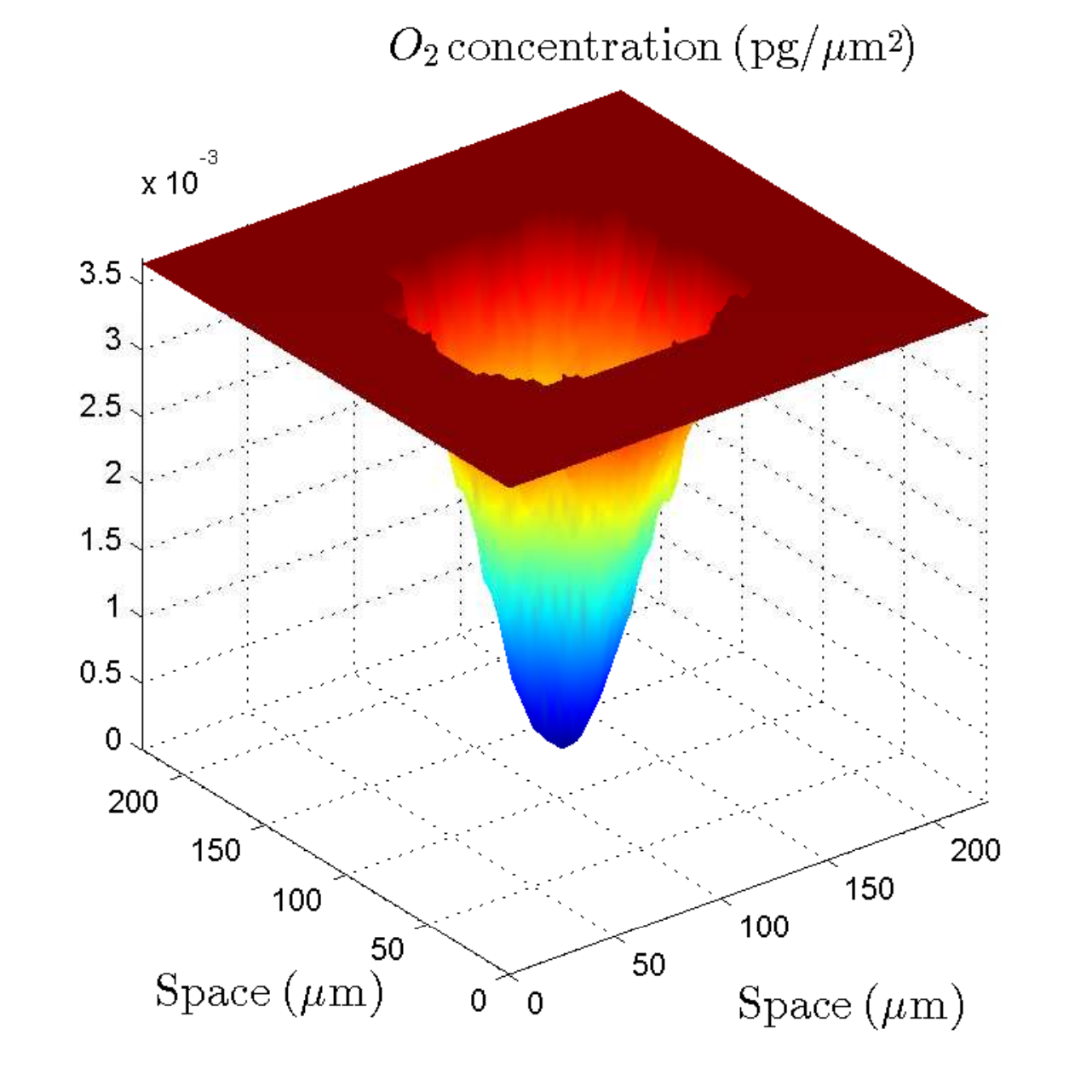}} \\\vspace{1 cm}
\subfigure[]{\includegraphics[scale=0.35, angle=0]{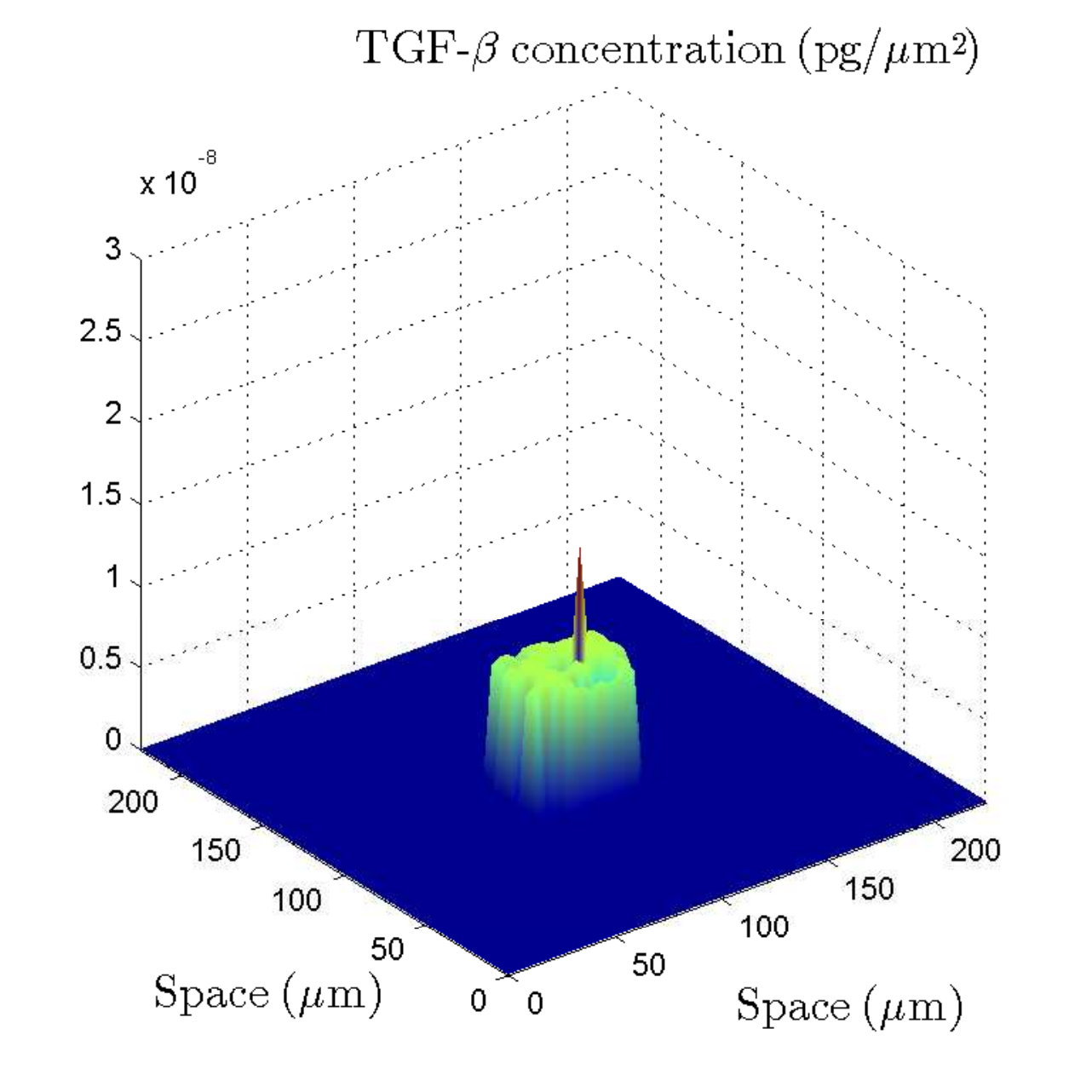}}  
\subfigure[]{\includegraphics[scale=0.35, angle=0]{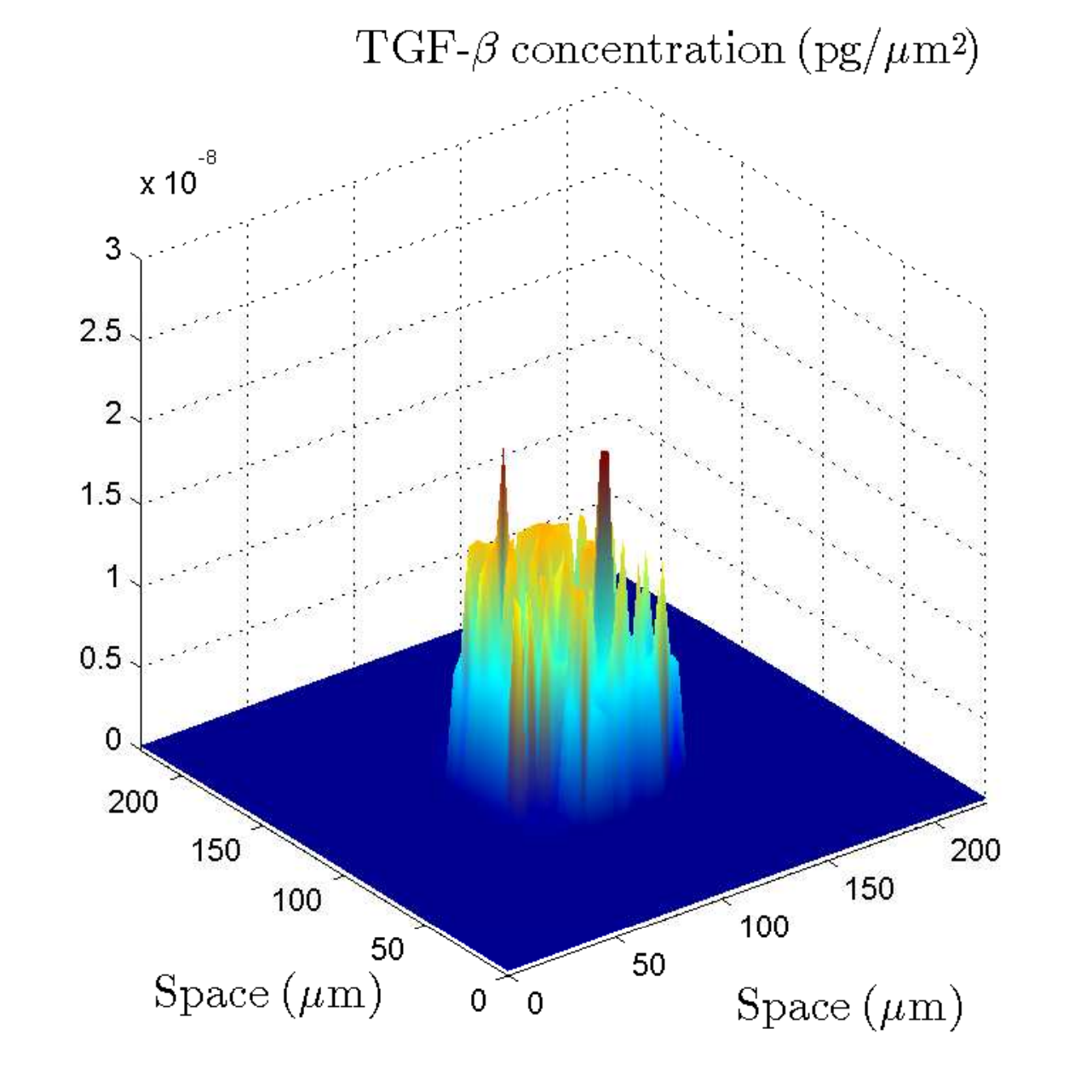}}  
\subfigure[]{\includegraphics[scale=0.35, angle=0]{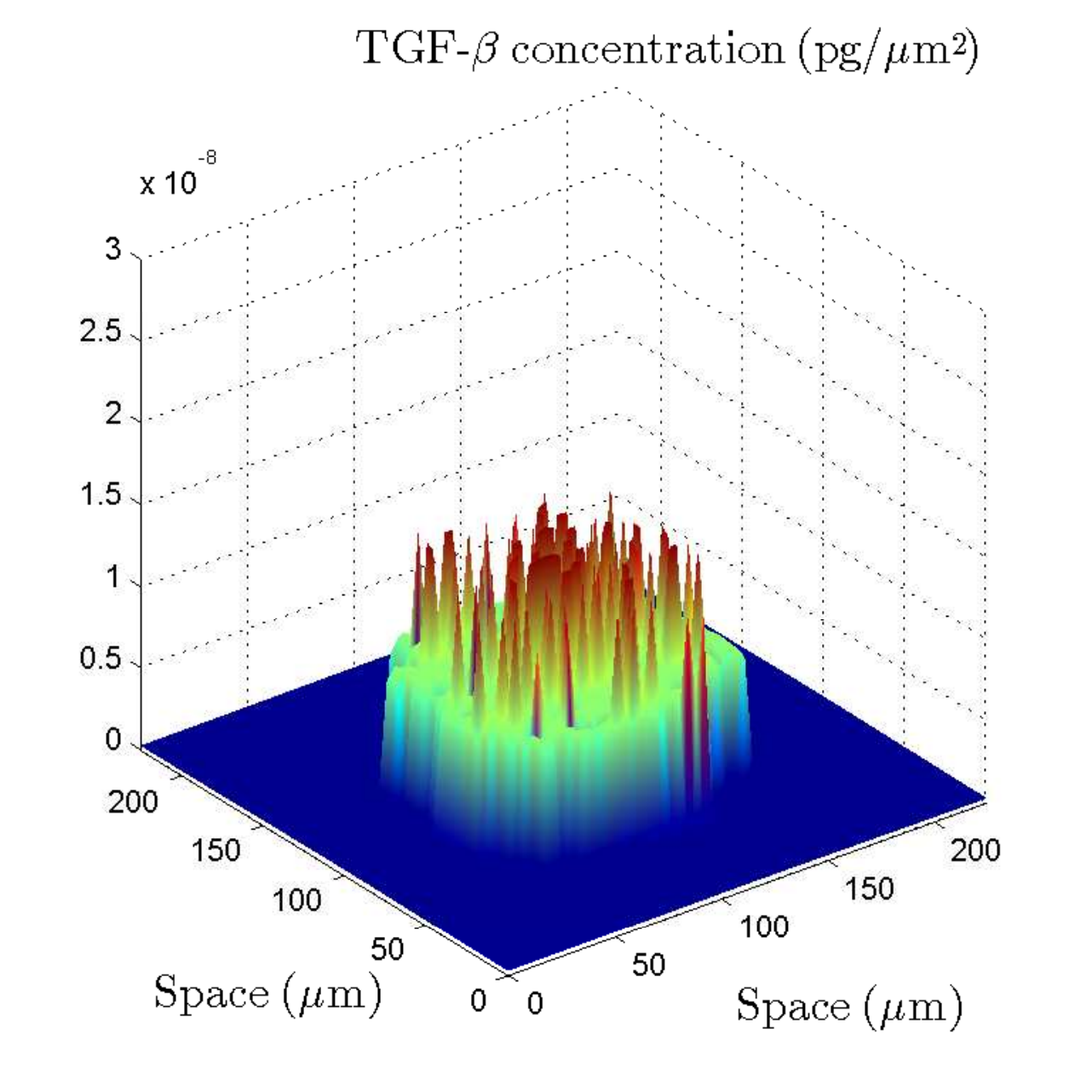}}
\caption{
Spatial distribution of the oxygen (a)--(c) and the TGF-$\beta$ (d)--(f) in the case of 21\% oxygen environmental concentration, at $t=24, 48, 72$ h. 
Equations \eqref{eq:complete}$_{4,5}$ are solved in a domain $\Omega=\left[0,225\right]\times\left[0,225\right]$ ($\mu$m$^2$), with the model parameters given by Table~\ref{tab-param-dim}.}
\label{fig:oxygen_tgf_concentration}   
\end{figure}

First, we analyzed the normal culture condition which corresponds to the 21\% concentration of the oxygen $O_2$. The state of the CSp at the initial and three following times is presented at Figure~\ref{fig:sphere_21percent}. Differentiation levels are marked by different colors: green for the least differentiated cells ($\varphi=1$), blue color is for the intermediate level of differentiation ($\varphi=2$), while the red color labels cells with the highest degree of maturation ($\varphi=3$). The reported results are in good agreement with the biological observations. We reproduced the CSp biological system consisting of a central core of less differentiated but faster proliferating cells surrounded by more specialized ones. The oxygen and TGF-$\beta$ concentrations at different times are shown at Figure~\ref{fig:oxygen_tgf_concentration}.

\begin{figure}[htbp!]
\centering
\subfigure[]{\includegraphics[scale=0.5, angle=0]{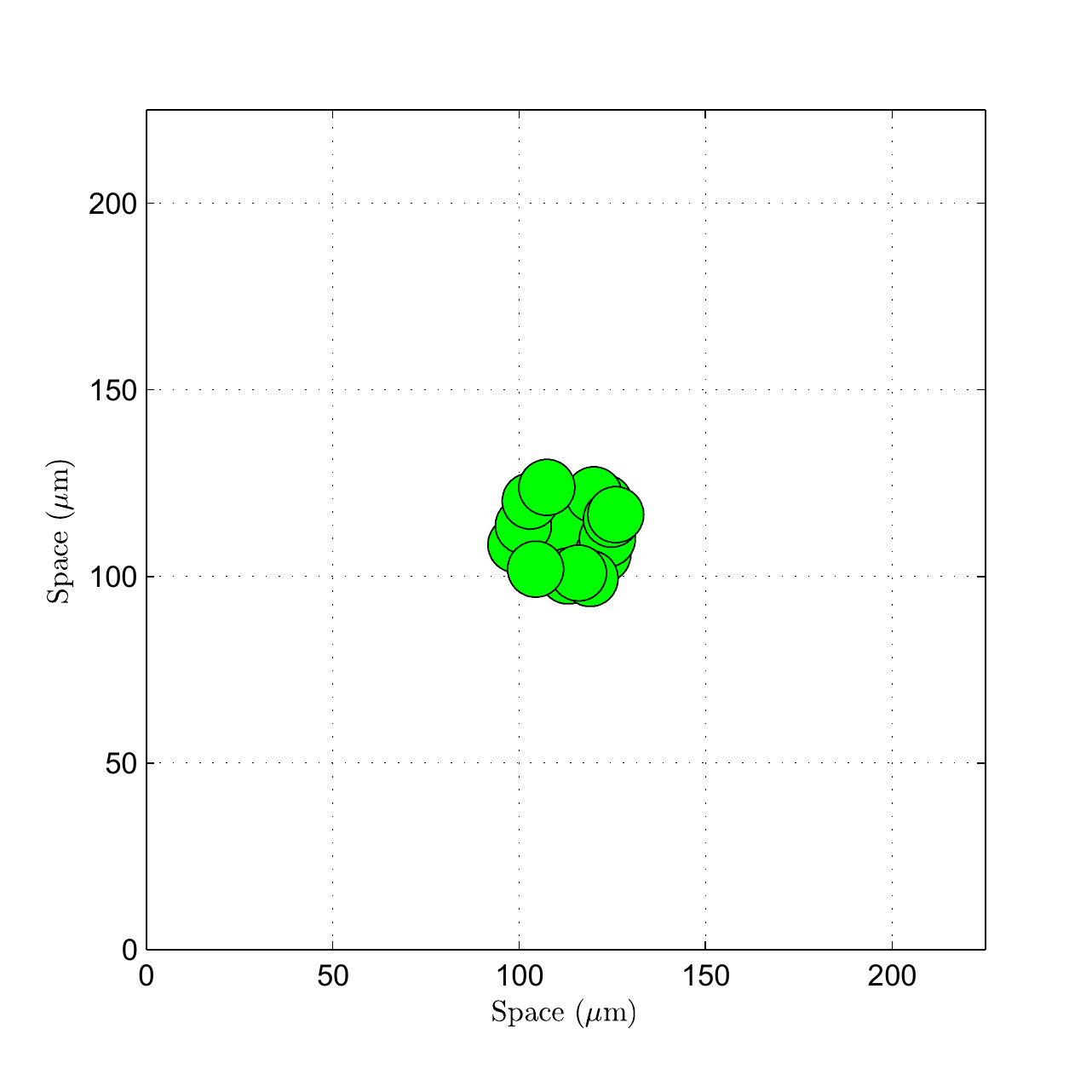}}  
\subfigure[]{\includegraphics[scale=0.5, angle=0]{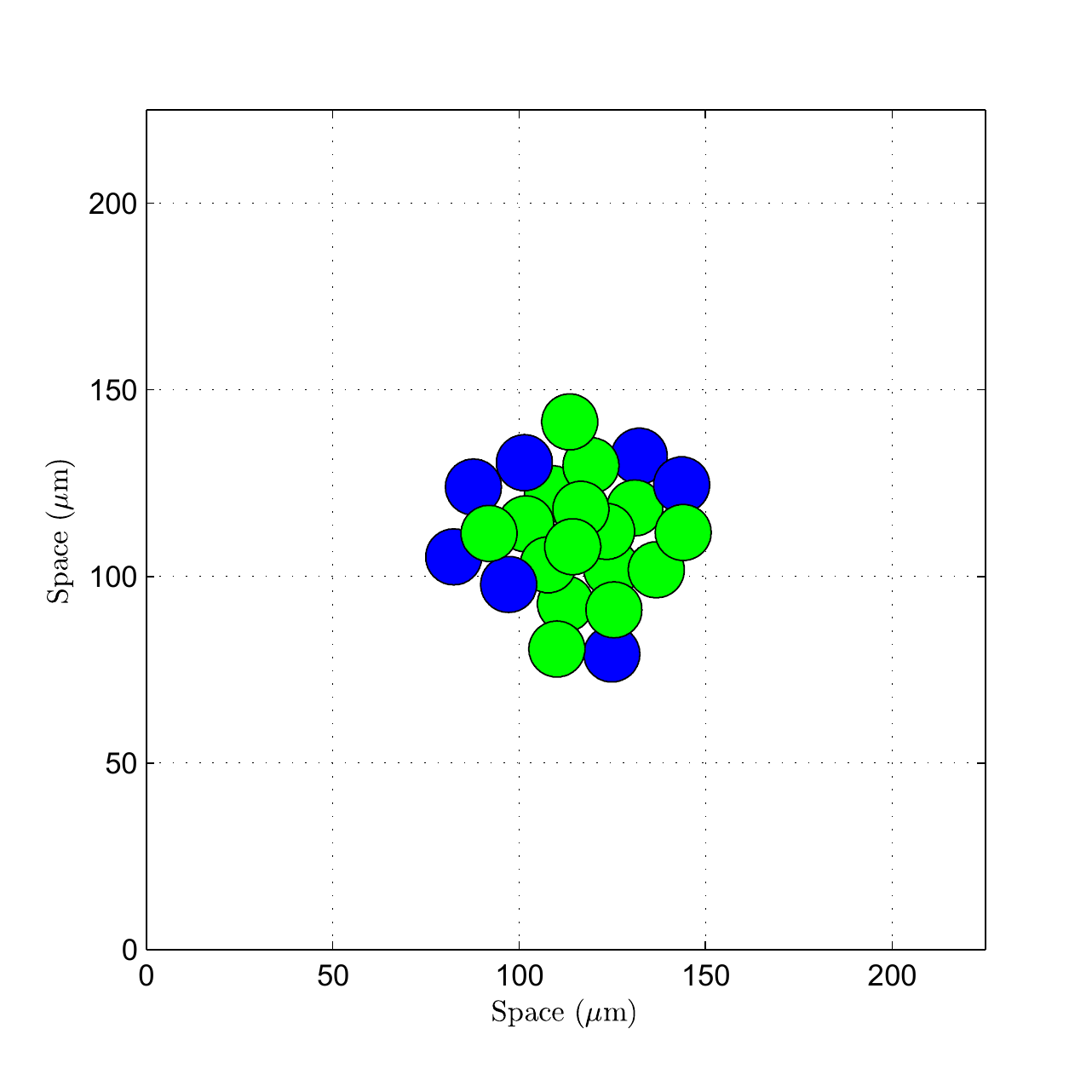}}  \\
{\scriptsize $t=0,  N_1=15, N_2=0,  N_3=0,  N_d=0$}  \qquad \qquad  
{\scriptsize $t= 24h,  N_1=16, N_2=7,  N_3=0,  N_d=0$} \\ \vspace{1cm} 
\subfigure[]{\includegraphics[scale=0.5, angle=0]{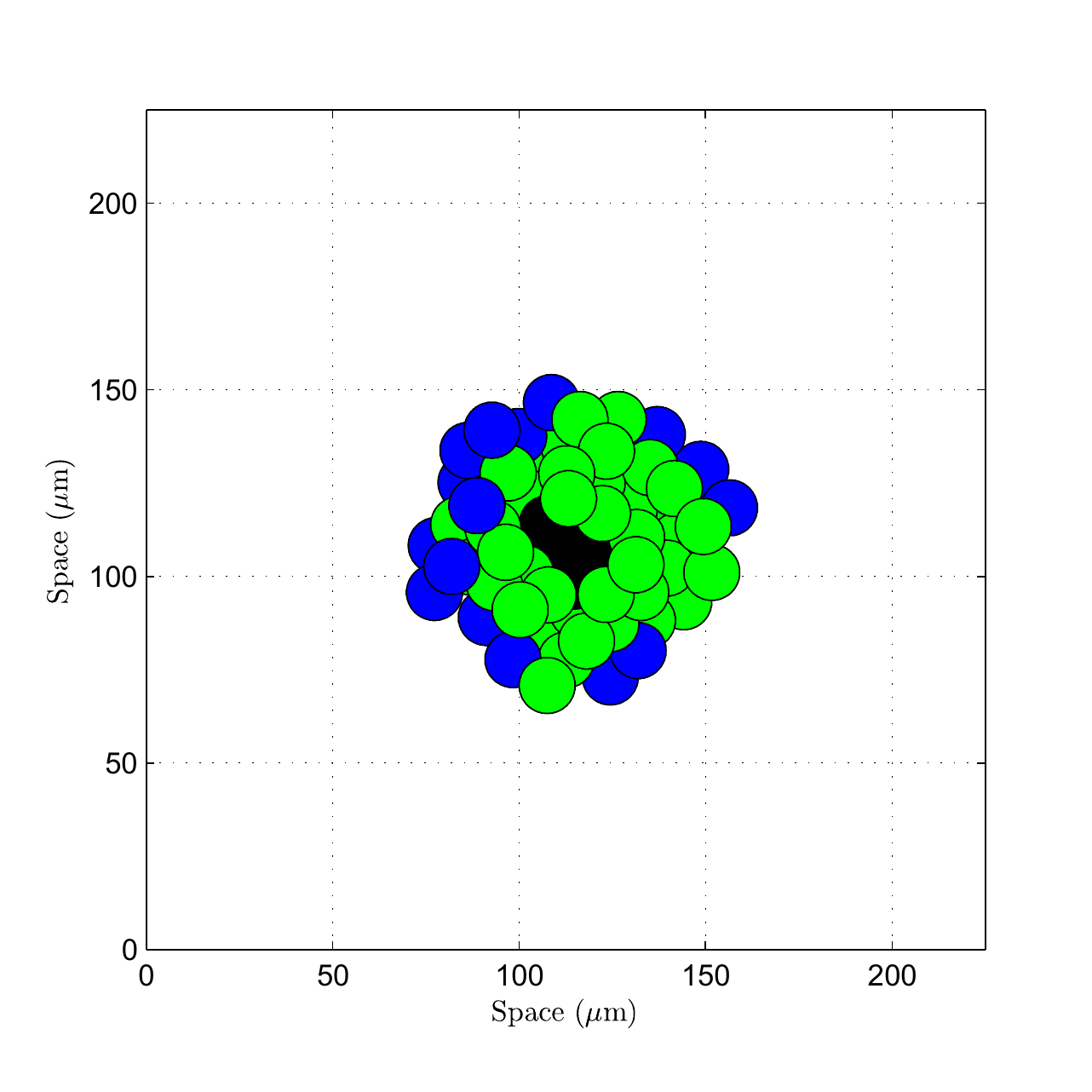}} 
\subfigure[]{\includegraphics[scale=0.5, angle=0]{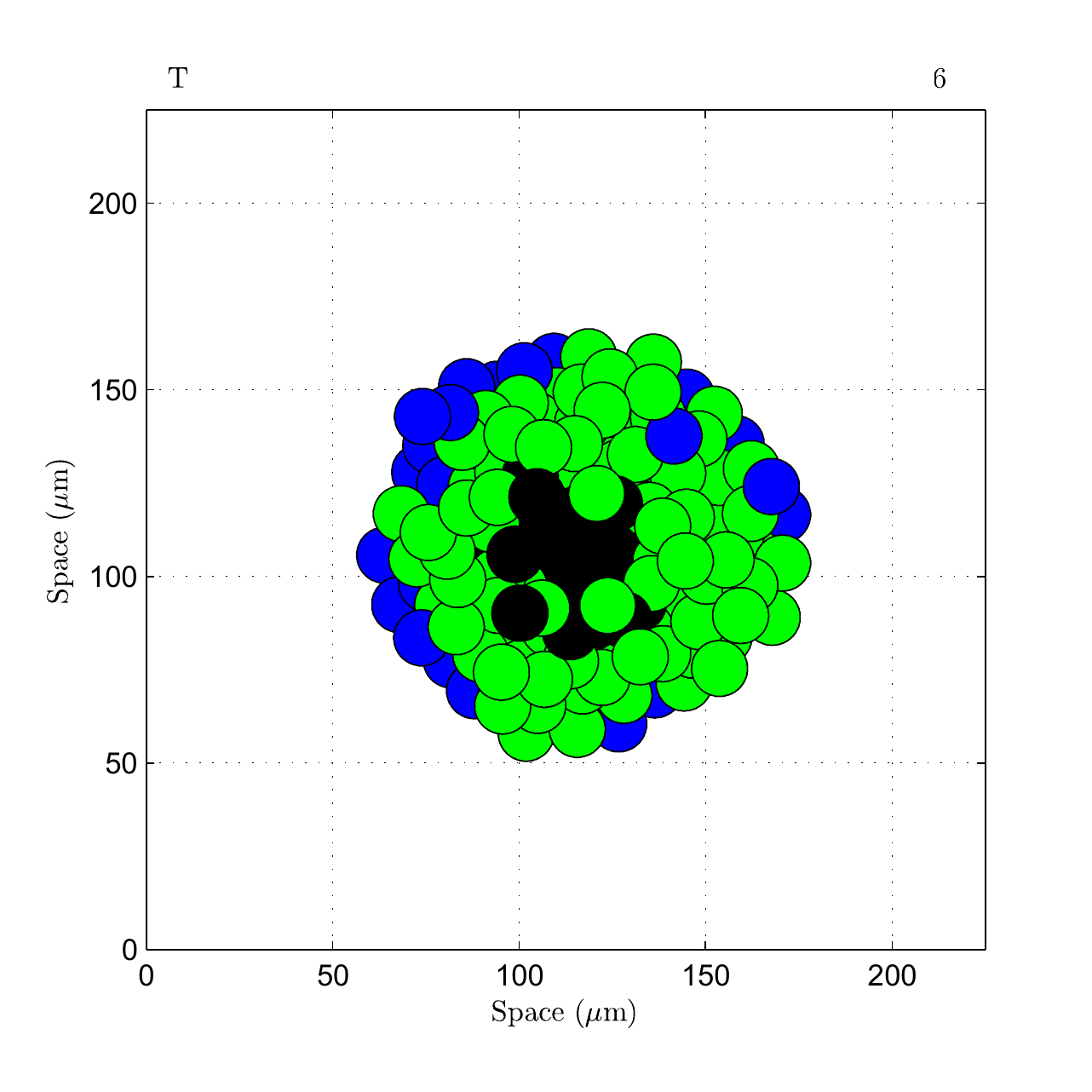}} \\
{\scriptsize $t=48,  N_1=41, N_2=16,  N_3=0,  N_d=7$}  \qquad \qquad  
{\scriptsize $t= 72h,  N_1=114, N_2=2,  N_3=0,  N_d=36$} \\ \vspace{1cm}  
\caption{
(a)--(d) Numerical simulation of the CSp growth for a $5\%$ oxygen environmental concentration, at times $t=0, 24, 48, 72$ h. System \eqref{eq:complete} is solved in a domain $\Omega=\left[0,225\right]\times\left[0,225\right]$ ($\mu$m$^2$), with the model parameters given by Table~\ref{tab-param-dim}. Differentiation levels are marked by different colors: green for the least differentiated cells ($\varphi=1$), blue color is for the intermediate level of differentiation ($\varphi=2$), while the red color labels cells with the highest degree of maturation ($\varphi=3$). Black region indicates a necrotic/apoptotic core composed by dead cells. As in Figure \ref{fig:sphere_21percent}, $N_1$, $N_2$, $N_3$, $N_d$, in the subplot captions show the composition of the CSp at the displayed times.}    
\label{fig:sphere_5percent}
\end{figure}

Then we considered 
the hypoxic culture conditions in the CSp environment, that is the $5 \%$  oxygen concentration. 
The corresponding structure and composition of CSp, shown at Figure~\ref{fig:sphere_5percent}, match the real situation. 
The size of the sphere is larger than that experimentally observed, because of the equally circular shape of cells and their constant radius in our model.

\begin{figure}[htbp!]
\centering
\subfigure[]{\includegraphics[width=0.5\textwidth]{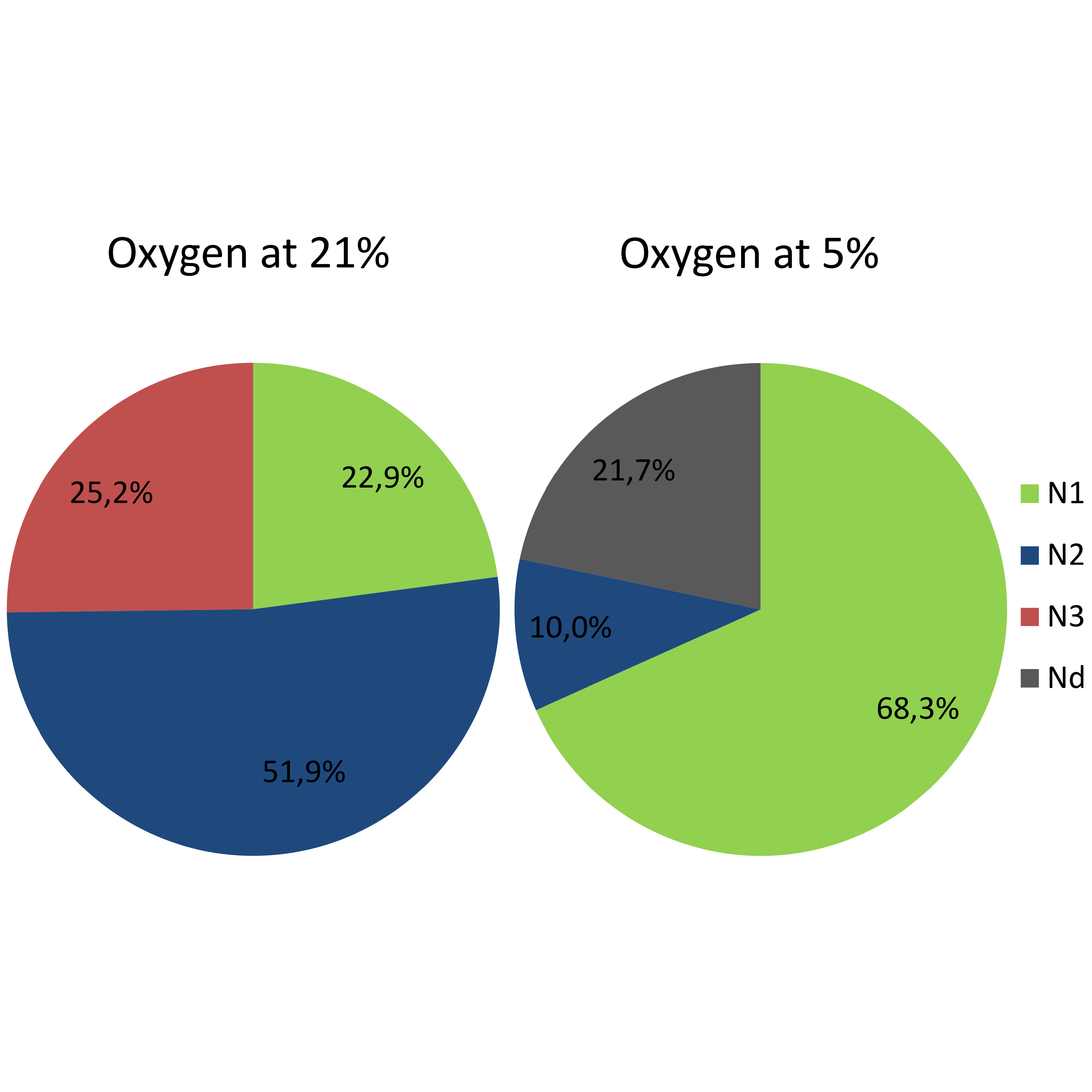}\label{fig:diameter_evolution_a}}  \hspace{-0.25 cm} 
\subfigure[]{\includegraphics[width=0.5\textwidth]{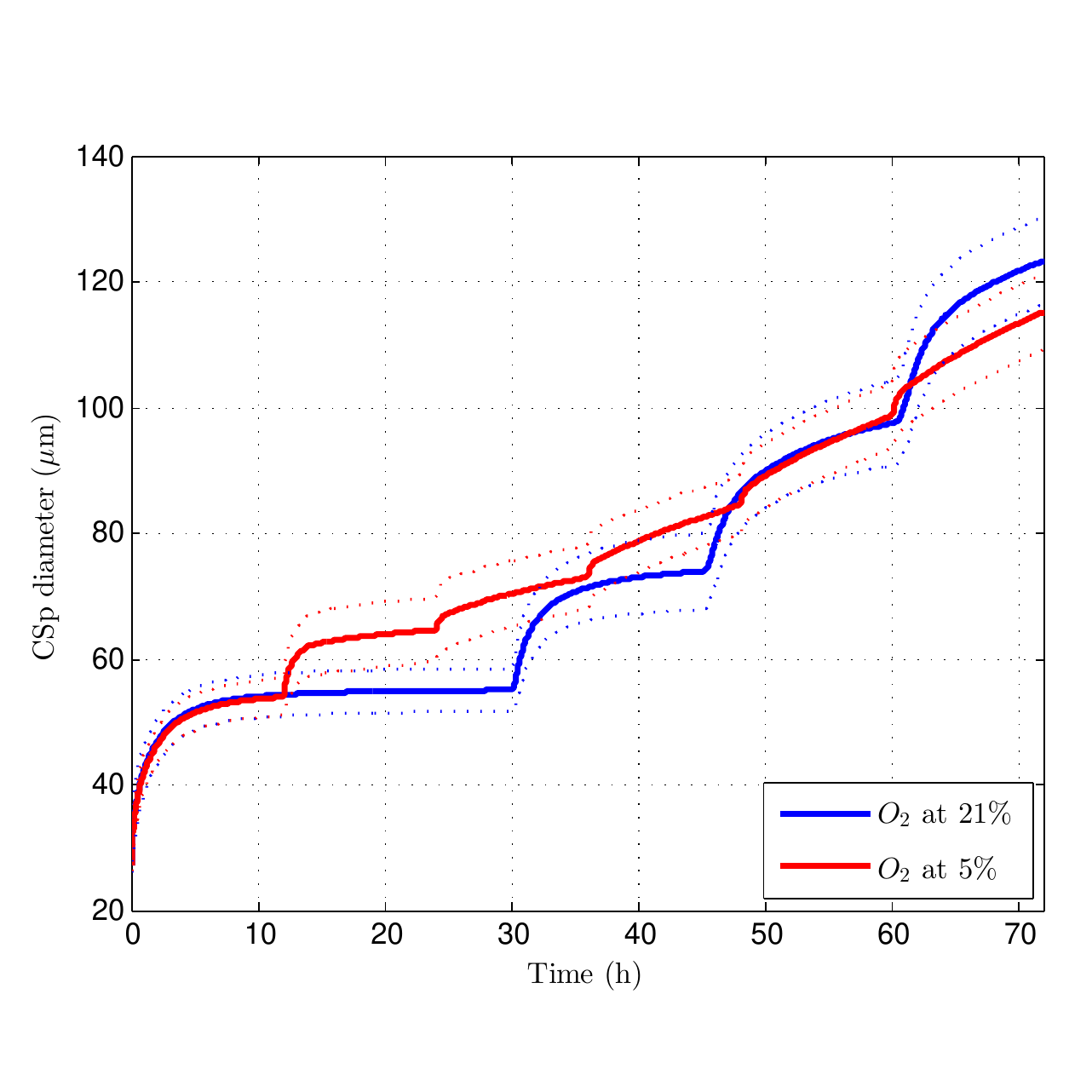}\label{fig:diameter_evolution_b}} \\ 
\subfigure[]{\includegraphics[width=0.5\textwidth]{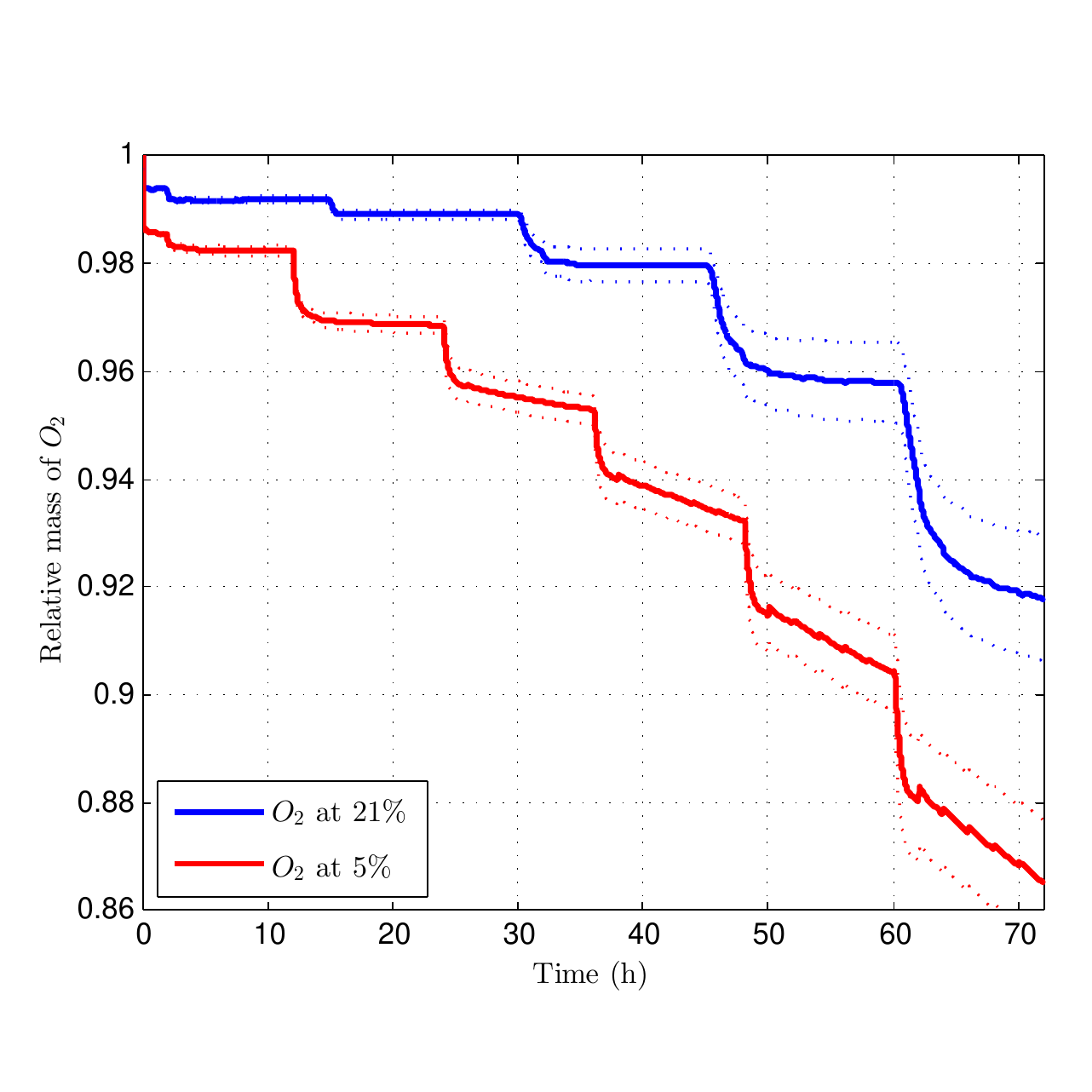}\label{fig:diameter_evolution_c}} \hspace{-0.25 cm}
\subfigure[]{\includegraphics[width=0.5\textwidth]{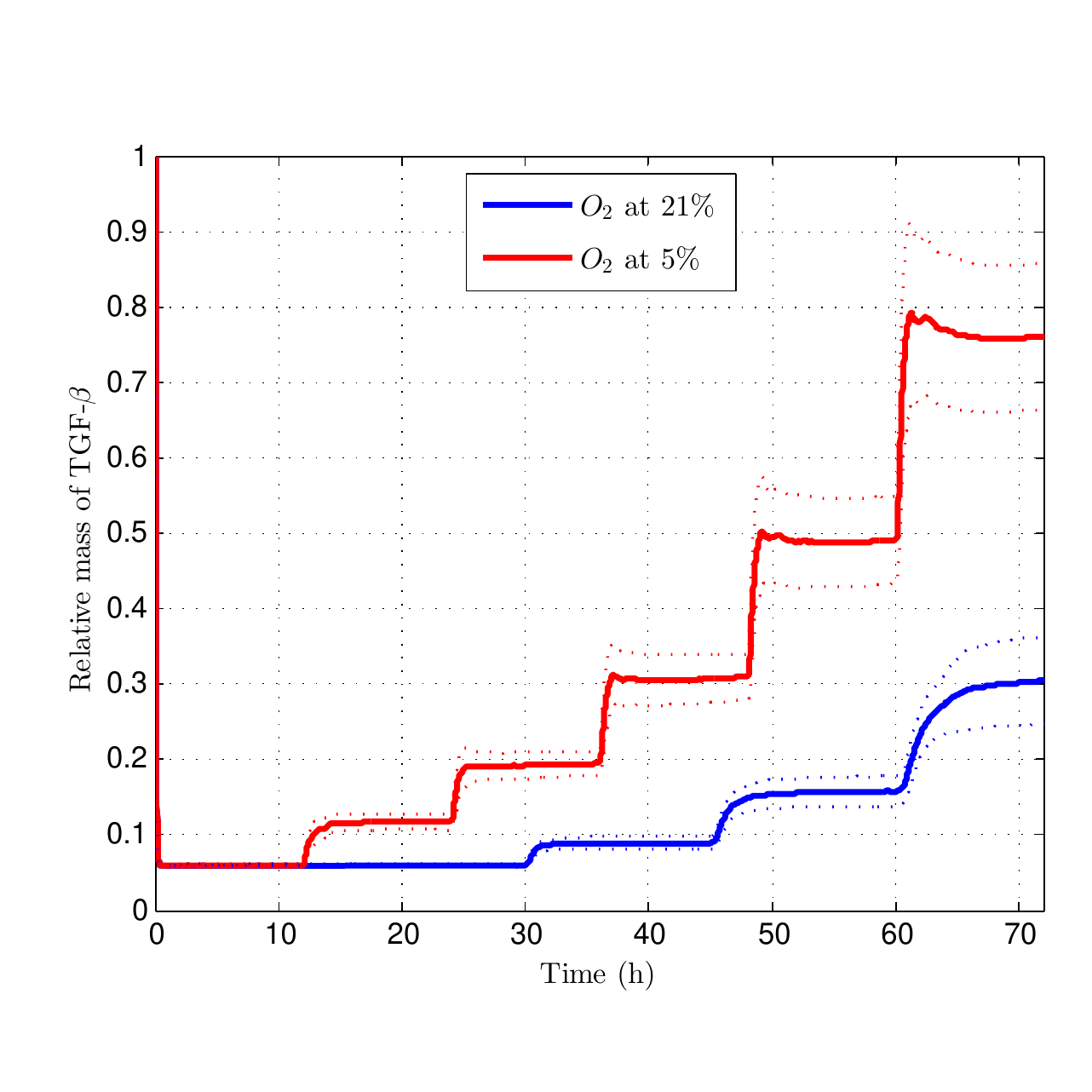}\label{fig:diameter_evolution_d}}
 \caption{(a) Pie charts describing the relative percentage distribution of the three types of cells (e.g. $(N_1/N_{\text{tot}})*100$) of a CSp in the case of $21$\% and $5$\% oxygen environmental concentration. The presented values are the averages over 100 independent simulations. (b) The diameter of the CSp as a function of time. (c)--(d) Oxygen and TGF-$\beta$ relative masses, $m_{\text{rel}}(t):=m(t)/m(0)$ versus time. For all the figures blue and red curves corresponds respectively to 21\% and 5\% oxygen concentration. In figures (b)--(d) the solid line represents the average value computed over 100 simulations, while the dotted curves show the interval corresponding to one standard deviation.}  
\end{figure}

To 
compare and summarize the above results we show at Figure~\ref{fig:diameter_evolution_a} in the form of pie charts the cell composition after 72 h in the two above cases. Similarly, Figure~\ref{fig:diameter_evolution_b} presents the growth of CSp diameter, and Figure~\ref{fig:diameter_evolution_c}--\ref{fig:diameter_evolution_d} the time evolution of the ratio between the total mass at time $t$ and the initial mass, $m_{\text{rel}}(t):=m(t)/m(0)$, respectively for oxygen and TGF-$\beta$. We observe that the oxygen levels decrease more at the hypoxic conditions (5$\%$) due to the increased proliferation and, as a result, initially the diameter of the sphere at $5$\% is larger than at $21$\%. However, the increasing deficiency of the oxygen lead to the formation of the necrotic core and the slower growth. Because of the initial condition equal for each cell and equal maturation time, cells undergo a higher degree of synchronisation. As a result we observe the characteristic steps in the curves describing the growth of the CSp diameter, in correspondence of the beginning of the proliferation cycle. 


The proposed model appears to be in 
a good agreement with the 
observed biological experiments at hand \citep{mess,chim1}. We were able to reproduce the basic structures of CSps at two different oxygen concentration levels. In case of $21$\% of the environmental concentration we observed the central region with proliferating, undifferentiated cells and the external ring of less multiplicating but differentiating cells. On the other hand, at $5$\% an evident necrotic core is observed, without many cells at higher differentiation level in the surrounding layers. 

\newpage
\section{Conclusions and perspectives}\label{conclusion}
In this paper we developed a hybrid mathematical model describing the dynamics of cardiac biopsy-derived stem cells leading to the formation of the so-called Cardiosphere from a cluster of a few cells.
 The complexity of the problem is extremely high and the biological phenomena leading to the formation of a CSp are still not completely understood. Moreover the quantitative characterization of the processes misses the estimate of some parameters essential for the accuracy of a mathematical model. Nevertheless, the proposed mathematical model appears in perfect agreement with the expected biological and pharmacological application shown in \citet{mess,chim1,chim,forte}. In particular, we were able to reproduce the basic structures of CSps assuming that only two elements are sufficient to describe the overall growth and maturation of the CSps: the oxygen as nutrient's emblem and the TGF-$\beta$ as the chemical differentiation signal representative. Our choice to select these two key regulators of the CSps growth and differentiation gives us many more chances to easily foresee the main possible cellular consequences following experimental biological or pharmacological modifications, with a strong impact in both research quality and cost. The layered growth of the CSps, with a central proliferating region, surrounded by differentiated and less proliferating cells, has been observed in the numerical experiments. In this regard, the biological technology has a high potential for therapeutic purposes and possible strategies for further differentiated cells are under investigation \citep{Lee2013,Chen2014}. Furthermore, in a possibly more general framework, our modeling tool could be applied to many other cellular spheroid culture systems, such as tumor cells spheroids, induced-progenitor cells, embryoid bodies-derived stem cells.

\bibliography{biblio_stem} 
\bibliographystyle{spbasic}

\end{document}